\RequirePackage{ifpdf}
\documentclass[20pt, letterpaper, hyper]{JHEP3}
 \pdfoutput=1

\usepackage[utf8]{inputenc}

\usepackage{epic,eepic}
\usepackage{amsmath,amssymb,amsfonts}

\usepackage{graphicx}
\usepackage{csquotes}
\title{Holographic Entanglement Entropy of Multiple Strips}

\author{Omer Ben-Ami${}^1$, Dean Carmi${}^1$, Jacob Sonnenschein${}^1$,
	\\
	${}^1$ \textit{Raymond and Beverly Sackler Faculty of Exact Sciences \\
		School of Physics and Astronomy \\
		Tel-Aviv University, Ramat-Aviv 69978, Israel}
	\\

	\texttt{e-mails}: \textsf{omerben@post.tau.ac.il, carmidea@post.tau.ac.il, cobi@post.tau.ac.il}
	
}

\abstract{
We study holographic entanglement entropy (HEE) of $m$ strips in various holographic theories. 
We prove that for $m$ strips with equal lengths and equal separations, there are only 2 bulk minimal surfaces. For backgrounds which contain also ``disconnected" surfaces, there are only 4 bulk minimal surfaces.
Depending on the length of the strips and separation between them, the HEE exhibits first order ``geometric" phase transitions between bulk minimal surfaces with different topologies. We study these different phases and display various phase diagrams. 
For confining geometries with $m$ strips, we find new classes of ``disconnected" bulk minimal surfaces, and the resulting phase diagrams have a rich structure. 
We also study the ``entanglement plateau" transition,  where we consider the BTZ black hole in global coordinates with 2 strips. It is found that there are 4 bulk minimal surfaces, and the resulting phase diagram is displayed. We perform a general perturbative analysis of the $m$-strip system: including perturbing the CFT and perturbing the length or separation of the strips.}

\preprint{TAUP-2988/14}
\begin{document}

\tableofcontents

\setcounter{footnote}{0}
\section{Introduction}

 Entanglement entropy (EE) is an important tool in studying quantum systems. It has applications in areas such as condensed matter and quantum gravity. Entanglement entropy is difficult to calculate in a QFT \cite{Bombelli:1986rw,Srednicki:1993im,Holzhey:1994we,Callan:1994py,Calabrese:2009qy,Casini:2009sr,Solodukhin:2011gn}, but for CFTs with holographic duals there is a simple geometric formula proposed by \cite{Ryu:2006bv, Ryu:2006ef} and later derived in \cite{Casini:2011kv, Lewkowycz:2013nqa}.

In this paper we study the holographic entanglement entropy (HEE) of multiple strips. In situations where there is more than one locally minimal Ryu-Takayanagi surface, the prescription is to choose the absolute minimal surface amongst them. There exist phase transitions between the topologically distinct bulk minimal Ryu-Takayanagi surfaces. As a first example, recall the ``geometric" phase transitions studied by Headrick \cite{Headrick:2010zt}.
He discussed HEE and mutual information for 2 strip regions in a CFT. A ``geometric phase transition" occurs between two topologically distinct bulk surfaces, when one changes the separation or length of the strips, see Fig.~\ref{confutK}.
The mutual information of two entangling regions $a$ and $b$ is defined as:
\begin{equation}
\label{eq:mutualinformation}
I(a,b)= S(a) +S(b) - S(a\cup b)
\end{equation}
where $S(a)$ and $S(b)$ are the EE of the regions $a$ and $b$ respectively. 
$I(a,b)$ is zero (non-zero) for the bulk surfaces in Fig.~\ref{confutK} left (right), and there is a first order phase transition. 

For $m$ strips, the mutual information can be defined in several ways. Two definitions are \cite{Alishahiha:2014jxa}:
\begin{eqnarray}
\label{eq:trew1}
\hat{I} = \sum_{i=1}^{m} S(a_i) - S(a_1 \cup a_2 \cdots \cup a_m)
\end{eqnarray}

\begin{eqnarray}
\tilde{I} = \sum_{i=1}^{m} S(a_i) - \sum_{i<j}^m S(a_i \cup a_j ) +  \sum_{i<j<k}^m S(a_i \cup a_j \cup a_k) + \ldots + (-1)^m S(a_1 \cup a_2 \cup \ldots \cup a_m)
\end{eqnarray}

For recent papers on holographic mutual information see \cite{Headrick:2010zt,Swingle:2010jz,Fischler:2012uv,Mukherjee:2014gia,Tonni:2010pv,Allais:2011ys,Balasubramanian:2011at,Alishahiha:2014jxa,Hayden:2011ag,Asplund:2013zba}. For works on the CFT side see \cite{Calabrese:2009ez,Calabrese:2010he,Cardy:2013nua,Hartman:2013mia,Faulkner:2013yia,Schnitzer:2014zva}. 

\begin{figure}[!h]
	\centering
	\includegraphics[width= 80mm]{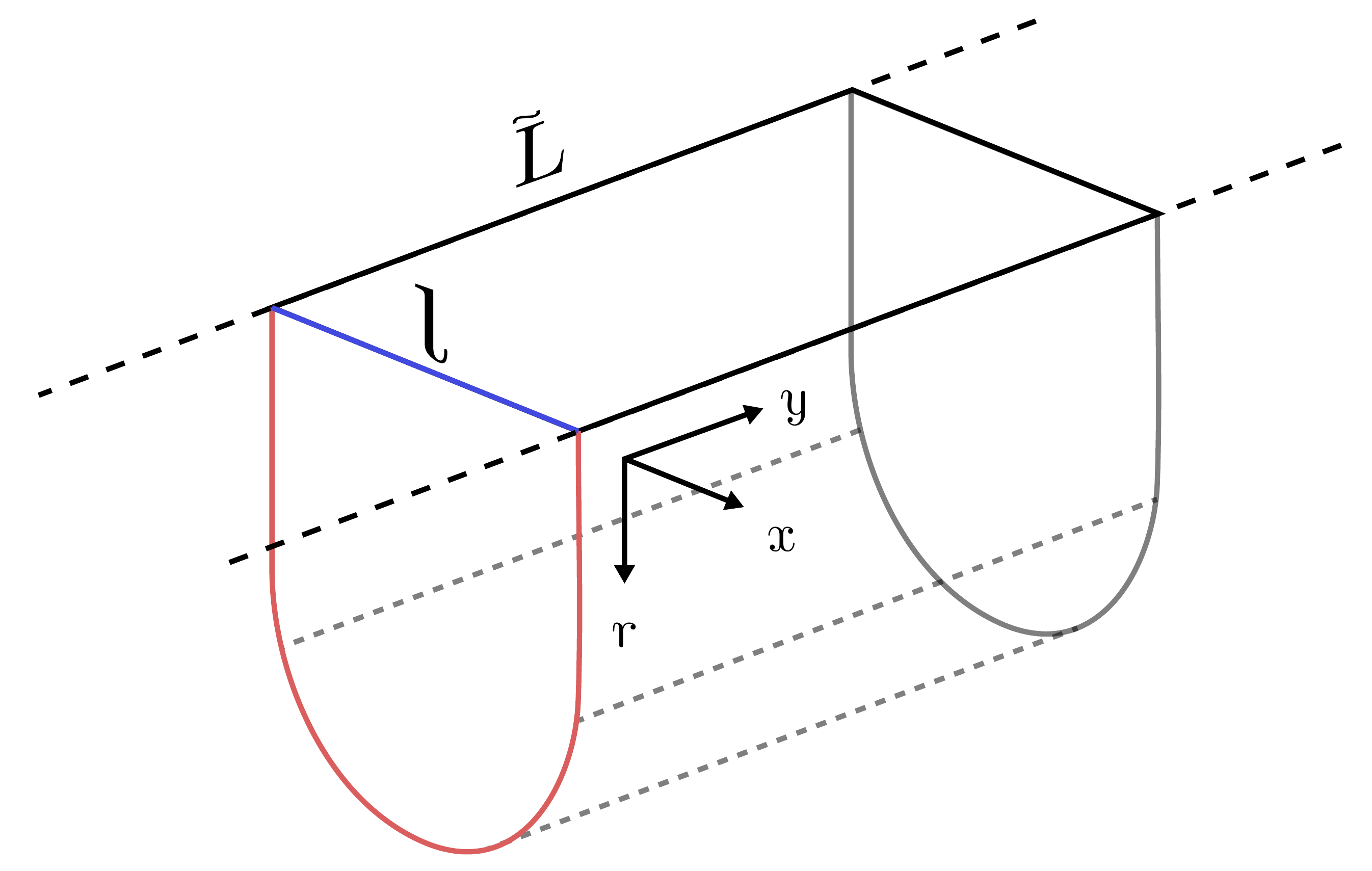}
	\caption{A strip configuration in 3 spacetime dimensions. The figure is in a constant time slice. $l$ is the length of the strip and $\tilde{L} \gg l$. Also shown in the picture is the bulk surface which extends in the direction $r$ and ends on the entanglement surface.\label{l1strip}}
\end{figure}

Holographic Entanglement entropy exhibits additional types of phase transitions. Using a generalization of the RT formula to non-conformal backgrounds, \cite{Klebanov:2007ws, Nishioka:2006gr} studied the entanglement entropy $S(l)$ for a single strip region (Fig.~\ref{l1strip}) of length $l$ in confining backgrounds. They discovered that a first order phase transition occurs between two distinct bulk surfaces (a ``connected" and a ``disconnected" surface) at a critical strip length $l=l_{crit}$, Figs.~\ref{confiningonestrip}, \ref{conf2}. This phase transition was conjectured to be related to the Hagedorn transition \cite{Klebanov:2007ws, Nishioka:2006gr} .

A third class of HEE transitions are the ``Entanglement-plateaux" transitions discussed in \cite{Hubeny:2013gta,Blanco:2013joa}. These transitions occur in holographic CFTs at finite temperature defined on a compact space. The simplest example is the BTZ black hole in global coordinates. In this example, consider an entangling region which is a segment of angle $\theta$ on the boundary circle, Fig.~\ref{btzonestrip}. Enlarging $\theta$, there is a phase transition at $\theta= \theta_{c}$ to a ``disjoint" Ryu-Takayanagi surface containing a piece which wraps the horizon of the black hole. 

In our work, we generalize the above phenomena to 2 or more strips. We plot phase diagrams which encode the different regions of the topologically distinct bulk minimal surfaces. These phase diagrams exhibit an interplay between the ``geometric phase transitions" and the other phase transitions mentioned above.

For CFTs with 2 or more strips, various phase diagrams are displayed. The effect of changing $m$ and $d$ will also be studied: reducing $m$ or $d$ causes disentanglement.

A similar analysis is done for holographic confining backgrounds with 2 or more strips. The $m$ strip HEE is characterized by a new class of ``disconnected" bulk minimal surfaces such as $S_C$ and $S_D$, Fig.~\ref{3stripsmu3}.
We display phase diagrams for different values of the number of strips $m$. In the limit $m \to \infty$, the area of the region $S_B$ in the phase diagram shrinks to zero. Using the definition Eq.\ref{eq:trew1} of the mutual information, the regions $S_A$ and $S_D$ have $\hat{I}^{(S_A)}= \hat{I}^{(S_D)}=0$, whereas $\hat{I}^{(S_C)}$ depends only on $x$ (and not $l$). Therefore, in the limit $m \to \infty$, $\hat{I}$ is independent of $l$ in all of the parameter space, unlike the CFT case which we discuss below in Eq~\ref{eq:wind1}.


The ``entanglement plateaux" transition is studied in the BTZ black hole geometry in global coordinates for 2-strips. The phase diagram corresponding to the 4 different bulk minimal surfaces is studied. When enlarging the radius of the black hole the phase diagram changes such that three of the regions shrink. For a very large black hole the parameter space is dominated by one of the phases.

Two additional examples of holographic backgrounds are discussed: A Dp-brane background dual to a non-CFT, and CFTs with temperature. We also discuss a ``correspondence" between holographic Wilson loops at finite $T$ and HEE for confining backgrounds, and vice versa. Using this, we argue that our results can be applied (qualitatively) to holographic Wilson loops.

We perform a more general perturbative analysis, and study how the HEE of $m$-strips changes when the CFT is perturbed. A ``positive" perturbation of the CFT, tends to break the ``joint" bulk surfaces into ``disjoint" ones. Conversely, a ``negative" perturbation will tend to join together bulk ``disjoint" surfaces.

We study how the HEE of $m$-strips changes when we perturb the separations and lengths of the strips. One result, is that a configuration with equal ``inner" separations is a maximum of the HEE with respect to perturbing these ``inner" separations.
 A second result is that enlarging the length of an ``inner" strip, reduces the HEE. This is opposite behavior compared to the effect of enlarging an ``outer" or ``disjoint" strip.\\

We will derive the following theorems which will be useful to us:
\begin{itemize}
\item 
For $m$ strips of equal lengths and equal separations, the ``connected"\footnote{``Disconnected" surfaces are bulk surfaces which have parts that terminate at the end of the bulk space. $S_C$ and $S_D$ are examples of ``disconnected" surfaces. ``Connected" surfaces are bulk surfaces which are not ``disconnected".} minimal bulk surface is either $S_A$ or $S_B$, see Fig.~\ref{3stripsmu3}.  
This means that ``disjoint" surfaces $S_P$ (exemplified in Fig.~\ref{sp}), are not the absolute minimal surfaces for any values of $x$ and $l$. This theorem greatly simplifies the problem of $m$ strips with equal lengths and equal separations, since one has to consider only 2 bulk surfaces, instead of at least $2^{m-1}$ bulk surfaces.

\item 
For $m$ strips of equal lengths and equal separations, the only possible ``disconnected" minimal bulk surfaces are $S_C$ or $S_D$, see Fig.~\ref{3stripsmu3}. 
This means that bulk surfaces such as $S_Q$ and $S_R$ of Fig.~\ref{fig:disjointdisconnected}
, are not the absolute minimal surfaces for all values of $x$ and $l$.

\item 
Combining the two results above gives:
For $m$ strips of equal lengths and equal separations there are only 4 possible bulk minimal surfaces: $S_A$, $S_B$, $S_C$, and $S_D$. See Fig.~\ref{3stripsmu3}.\\
\end{itemize}

This paper is organized as follows.
 In section \ref{sec:mutual1} we discuss the HEE for CFTs with $m$-strips.
In section \ref{sec:book5} we discuss the HEE for holographic confining backgrounds with $m$-strips.
In section~\ref{sec:btz23} we study the ``entanglement plateaux" transition for 2-strips in the BTZ background.
In section \ref{sec:book9} we mention two additional examples of holographic backgrounds.
In section~\ref{sec:the2} we perform a more general perturbative analysis of the $m$-strips system.
In section~\ref{sec:theorems4} we exclude certain classes of bulk minimal surfaces.
In section~\ref{sec:discussion2} we discuss our results and future directions.
In Appendix~\ref{sec:BB5} we discuss holographic Wilson loops and its relation to holographic entanglement entropy.

\section{Phases of HEE in CFTs with multiple strips}
\label{sec:mutual1}

In a holographic theory dual to a  CFT, the Ryu-Takayanagi formula for a single strip region (Fig.~\ref{l1strip}) gives \cite{Ryu:2006bv, Ryu:2006ef,Nishioka:2009un}:
\begin{eqnarray}
	S_1(l) = \frac{1}{4G_N^{d+1}} \Bigg[ \frac{2R^{d-1}}{d-2}\bigg( \frac{\tilde{L}}{\epsilon} \bigg)^{d-2} -\frac{2^{d-1}\pi^{\frac{d-1}{2}}R^{d-1}}{d-2}  
	\bigg( \frac{\Gamma (\frac{d}{2d-2})}{\Gamma{(\frac{1}{2d-2})}}\bigg)^{d-1} \bigg( \frac{\tilde{L}}{l} \bigg)^{d-2}\Bigg]
\end{eqnarray}
where $d$ is the number of spacetime dimensions, $\epsilon$ is the UV cutoff, $R$ is the AdS radius, $l$ is the length of the strip, and $\tilde{L}\gg l$.  We use the notation $S_1$ to indicate the entanglement entropy of one strip. The first term is the ``Area law" and the second term is a finite term.

For our purposes the divergent term will not play a role (we will always ask questions about differences of entanglement entropies), and neither will the factor multiplying the finite term (which will always be just an overall factor which we set to be $1$). With this in mind, we simply write the finite term (or the log term in $2d$) as:
\begin{align}
	\label{eq:book1}
	S_1(l)&= - \frac{1}{l^{d-2}}\quad && d>2\\
	S_1(l)&= \frac{c}{3}\log (l/\epsilon) \quad && d=2
\end{align}
Notice that for a CFT the finite term has a closed analytic form (this is not true in general).

\subsection{CFT with two strips of equal length $l$}

\begin{figure}
\centering
\includegraphics[width= 160mm]{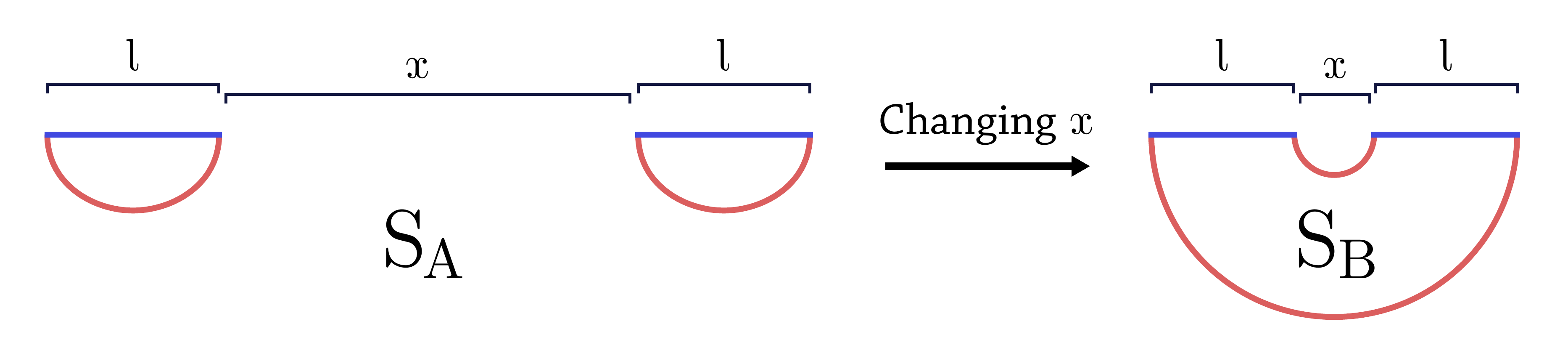}
\caption{ Illustration of the two minimal surfaces $S_A$ and $S_B$ for 2 strips in a CFT. There is a transition between the two surfaces when $x/l =f(d)$, where $f(d)$ depends only on $d$.\label{confutK}}
\end{figure}

Now consider two strips of equal length $l$ and separated by a distance $x$. As shown in Fig.~\ref{confutK}, we have two competing minimal surfaces \cite{Headrick:2010zt} $S_A$ and $S_B$ (which are ``joint" and ``disjoint" configurations). $S_B$ and $S_A$ indicate the areas of these ``joint" and ``disjoint" surfaces respectively. 
 
Using the translational symmetry of the 2 strip configuration, we can calculate $S_A$ and $S_B$ in terms of  the 1-strip result $S_1$ (this is obvious from Fig.~\ref{confutK}), the finite terms are:
\begin{align}
S_{A}(x,l)&= 2S_1(l)=  -\frac{2}{l^{d-2}}\\
S_{B}(x,l)&=S_1(2l+x)+S_1(x)= \ -  \frac{1}{(2l+x)^{d-2}} -  \frac{1}{x^{d-2}}  
\end{align}

\begin{figure}
\centering
\includegraphics[width= 80mm]{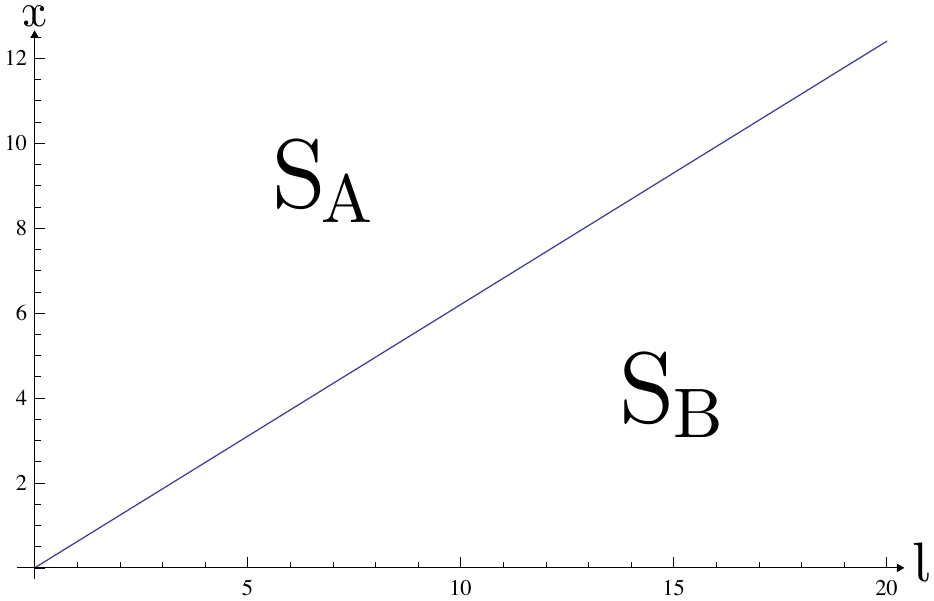}
\caption[l1strip]{The phase diagram for two strips in a $d=4$ CFT (dual to $AdS_5$). The corresponding minimal surfaces $S_A$ and $S_B$ are illustrated in Fig.~\ref{confutK} The transition line is the straight line $x/l=f(d)$}
\label{phasecftK}
\end{figure}

When $S_{A}=S_{B}$ there will be a transition between the two surfaces. This happens when
\begin{eqnarray}
\label{eq:qq4}
\frac{1}{(2+y)^{d-2}} +  \frac{1}{y^{d-2}}  = 2
\end{eqnarray}
where $y\equiv x/l$. The solution to this equation is $y_c=f(d)$, where $f(d)$ depends only on $d$.
The phase diagram in the $x-l$ plane is shown in Fig.~\ref{phasecftK} . It consists of the straight transition line $x=f(d)\cdot l$ which separates  the 2 phases. This is, of course, a consequence of the fact that there is no scale in a CFT, therefore $x/l$ can only depend on constants.
From Eq.~\ref{eq:mutualinformation}, the mutual information is zero for $S_A$ and non-zero for $S_B$\footnote{The divergent parts will always drop from the mutual information so we can simply use the finite terms}:
\begin{align}
\label{eq:wind1}
I_A(x,l)&= 0 
\nonumber\\
I_B(x,l)&= -\frac{2}{l^{d-2}} +  \frac{1}{(2l+x)^{d-2}} + \frac{1}{x^{d-2}}   
\end{align}

\subsection{CFT with $m$ strips}
\begin{figure}
	\centering
	\begin{minipage}{0.45\textwidth}
		\centering
		
		\includegraphics[width= 80mm]{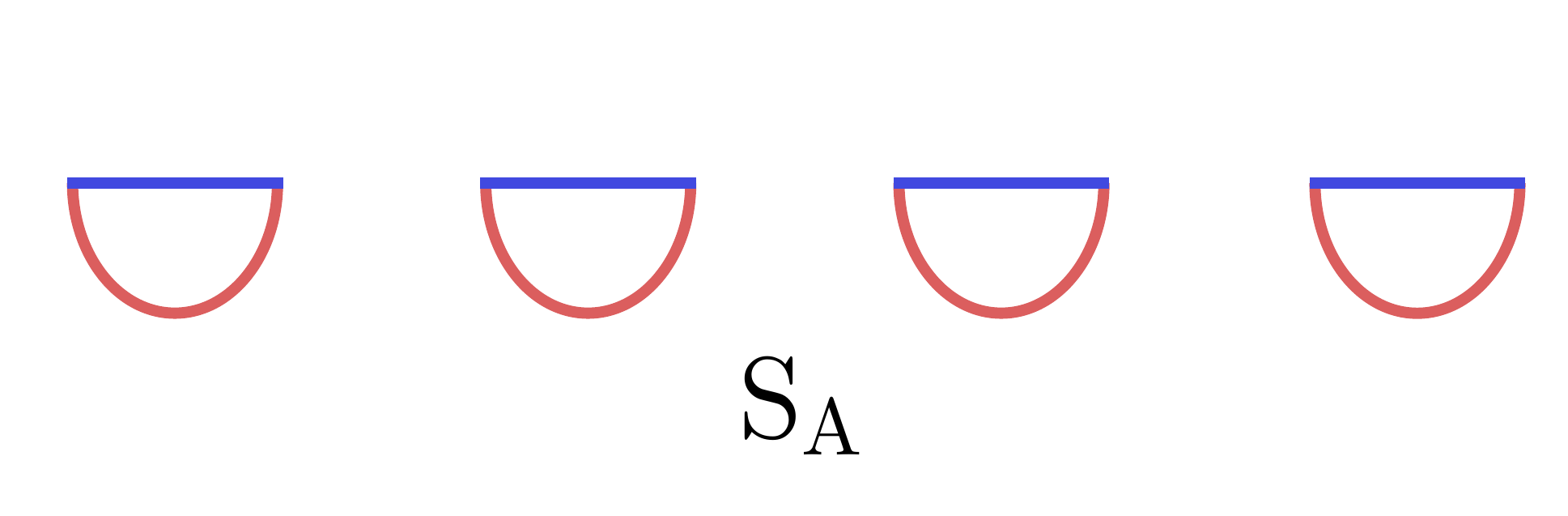}
		
	\end{minipage}\hfill
	\begin{minipage}{0.45\textwidth}
		\centering
		
		\includegraphics[width= 80mm]{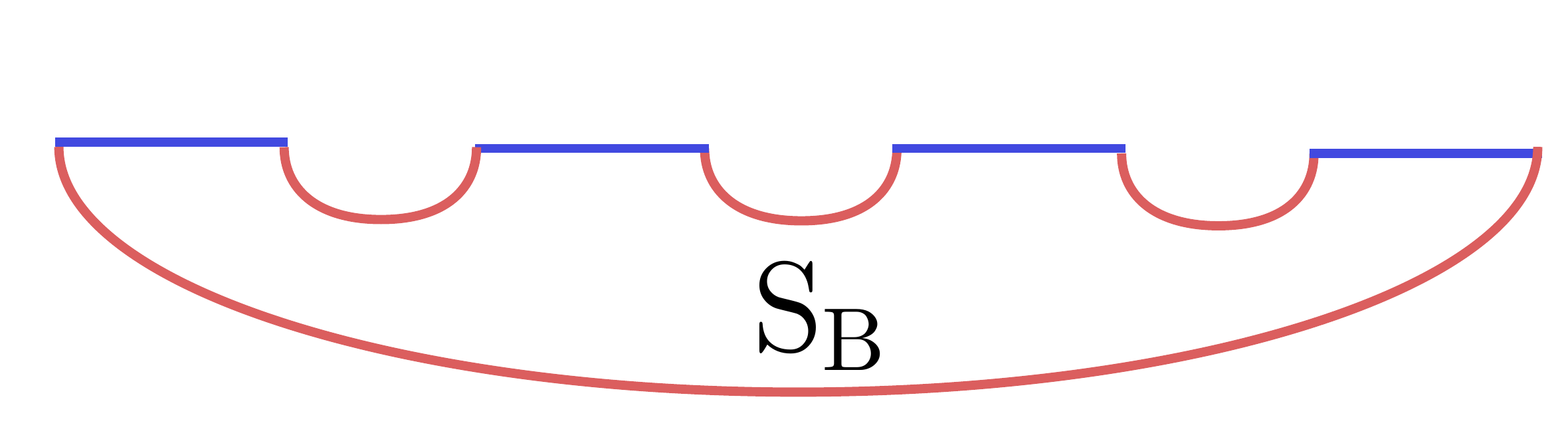}
	\end{minipage}
	
	\caption{Illustrating the surfaces $S_A$ and $S_B$ which are defined for arbitrary $m$. $S_B$ is defined to be the completely ``joint" surface, and $S_A$ is the surface for which all strips are ``disjoint". The plot illustrates $S_A$ and $S_B$ for the 4 strip case.\label{fig:multiple_sa_sb}}
\end{figure}
 We now consider the generalization to $m$ strips, see also \cite{Alishahiha:2014jxa}. For 2 strips we had 2 bulk minimal surfaces $S_A$ and $S_B$, Fig.~\ref{confutK}. For arbitrary $m$, we define $S_B$ to be the completely ``joint" surface, and $S_A$ is the surface for which all strips are ``disjoint". This is illustrated in Fig.~\ref{fig:multiple_sa_sb} for the case $m=4$. It is easy to write down the expression for $S_A$ and $S_B$ in terms of the 1-strip result $S_1$:
\begin{align}
\label{eq:qq2}
S_A(l_i,x_i)&= \sum_{i=1}^{m} S_1(l_i)\\
\label{eq:qq3}
S_B(l_i,x_i)&= \sum_{i=1}^{m-1} S_1\big(x_i) + S_1(\sum_{i=1}^{m-1} x_i+\sum_{i=1}^{m} l_i\big)
\end{align}
These equations will apply to arbitrary theories (with their corresponding finite term $S_1$), for arbitrary strip lengths $l_i$, and separations $x_i$.

Let us focus on the case of equal length strips and equal separations between them: $l_i=l$,and $x_i=x$. In  section~\ref{sec:theorems4} it is shown that $S_A$ and $S_B$ are the only bulk minimal surfaces for any $x$, $l$, $d$, and $m$.

Now consider a CFT with: $S_1(l)=-\frac{1}{l^{d-2}}$ (see Eq.~\ref{eq:book1}).
For equal lengths and equal separations, Eqs.~\ref{eq:qq2} - \ref{eq:qq3} become:
\begin{align} 
\label{eq:book11}
S_{A}(x,l)&= mS_1(l) = - \frac{m}{l^{d-2}}\\
\label{eq:book12}
S_{B}(x,l)&= (m-1)S_1(x)+S_1\big((m-1)x+ml\big) = -  \frac{m-1}{x^{d-2}} - \frac{1}{(ml+(m-1)x)^{d-2}} 
\end{align}

\begin{figure}
	\centering
	\begin{minipage}{0.45\textwidth}
		\centering

\includegraphics[width= 80mm]{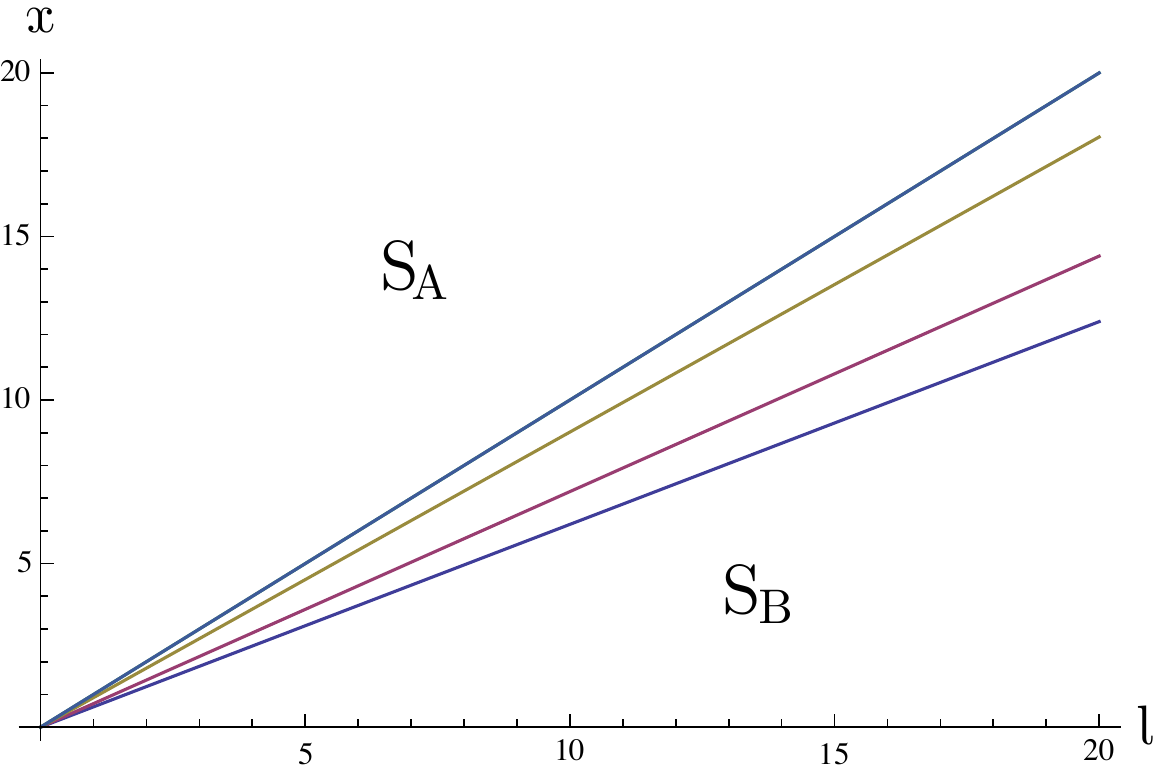}

	\end{minipage}\hfill
	\begin{minipage}{0.45\textwidth}
		\centering

\includegraphics[width= 80mm]{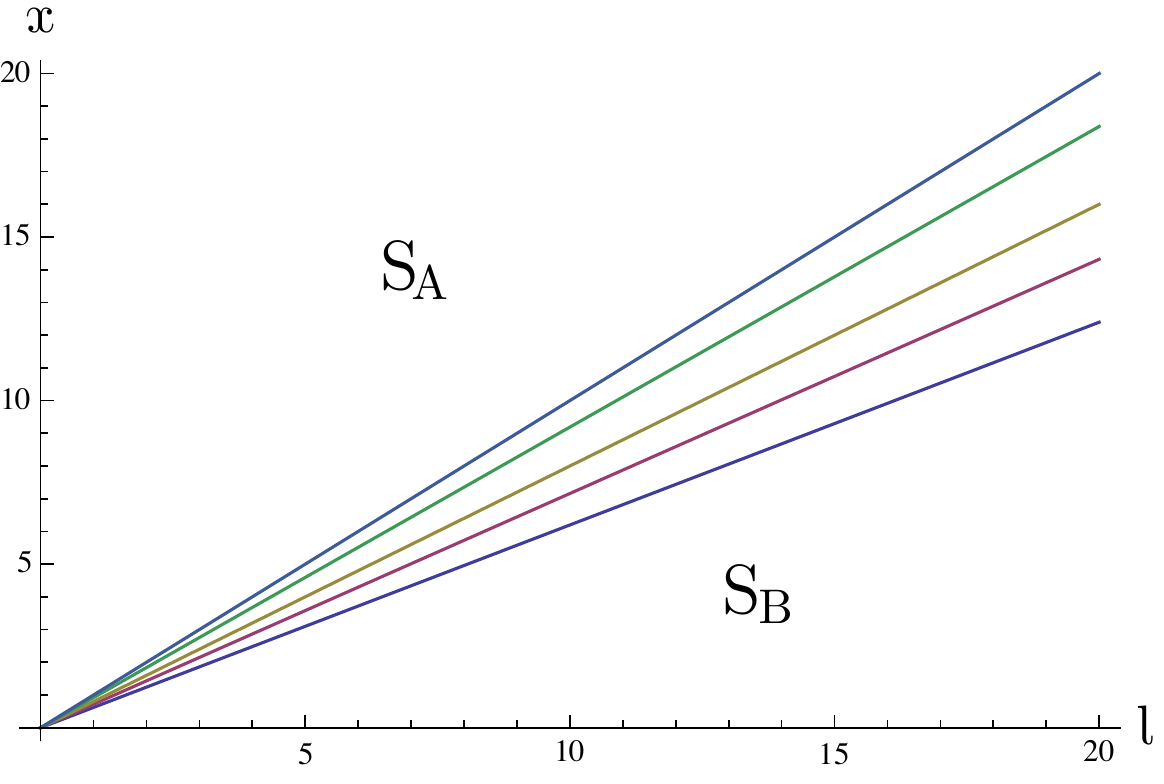}
	\end{minipage}

\caption{The phase diagram for a CFT and $m$-strips. The straight lines are the transition lines. \textbf{ Left:} $d=3$ and changing $m=2, 3, 10, 1000$. In the large $m$ limit the transition line has a slope of 1. \textbf{ Right:} $m=2$ and changing $d=3, 4,5, 10, 1000$. In the large $d$ limit the transition line has a slope of 1.\label{md9}}
\end{figure}

When $S_{A}=S_{B}$ there will be a transition between the two surfaces (See Eq.~\ref{eq:qq4} for the $m=2$ case). This happens when:
\begin{eqnarray}
\label{eq:qq1}
\frac{1}{(m+(m-1)y)^{d-2}} +  \frac{m-1}{y^{d-2}}  = m
\end{eqnarray}
where $y\equiv x/l$. The solution to this equation is $y_c=f_1(d,m)$, where $f_1(d,m)$ is a function of $d$ and $m$. The phase diagram in the $x-l$ plane consists of a straight transition line $x=f_1(d,m)\cdot l$.  For 3 and 4 dimensions we can solve this equation analytically:
\begin{align}
y_c&=\frac{\sqrt{1+m^2}-1}{m}& d=3\\
y_c&=\frac{\sqrt{m^2-m+1}-1}{m-1}& d=4
\end{align}
Fig.~\ref{md9} shows the phase diagram for different values of $m$ and $d$. If we make $m$ or $d$ larger, then the slope of the transition line will become larger. In the limit $m \to \infty$ or $d \to \infty$, the slope of the transition line approaches 1. Therefore reducing $d$ or $m$ (with constant $l$ and $x$) causes disentanglement.


\subsection{CFT with 2 strips of unequal length}
\begin{figure}
	\centering
		\begin{minipage}{0.45\textwidth}
			\includegraphics[width= 80mm,height=60mm]{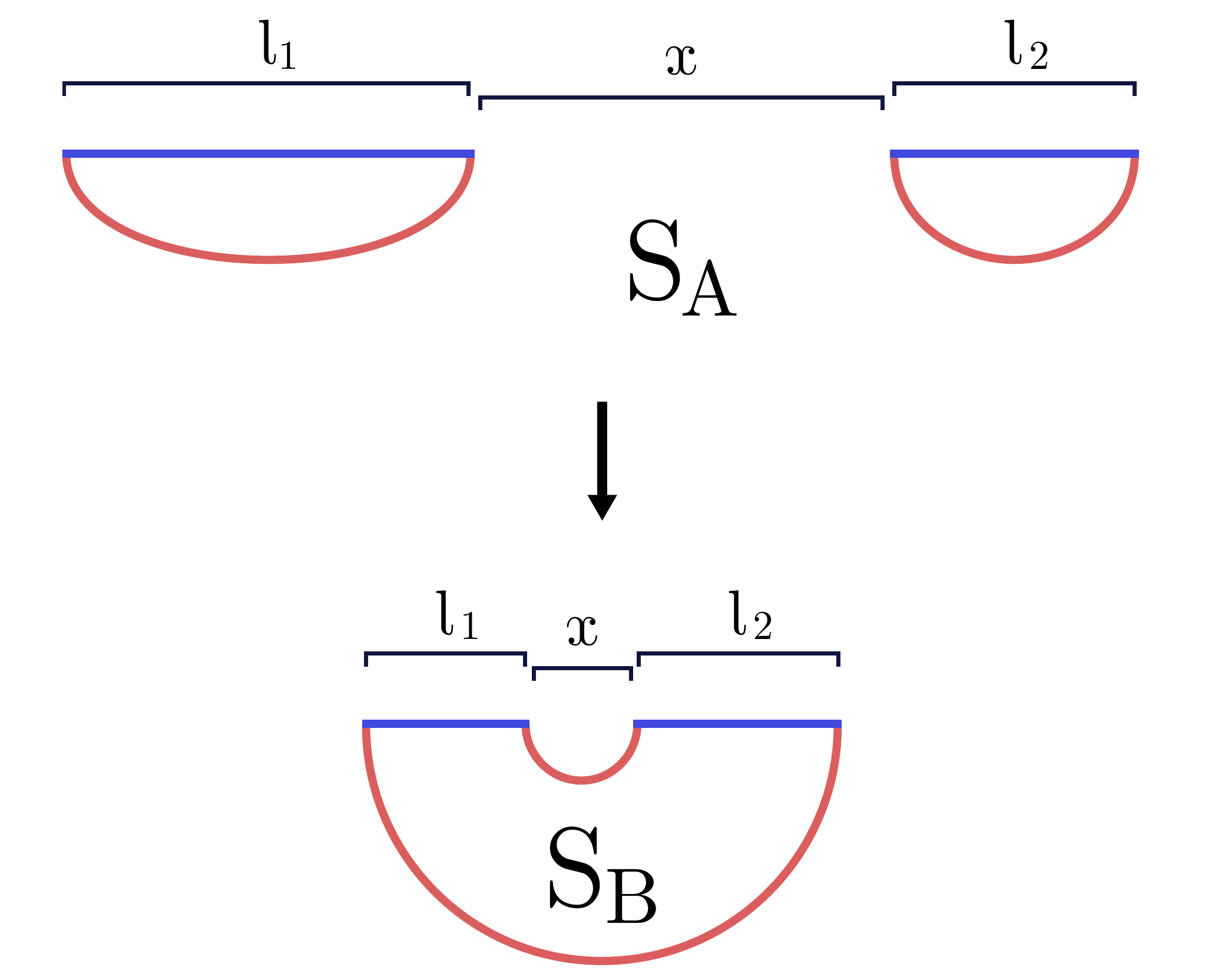}
			\caption{Illustration of the two bulk surfaces $S_A$ and $S_B$ for 2 strips of unequal lengths $l_1$ and $l_2$.\label{cft2stripsunequalseparation}}
		\end{minipage}\hfill
	\begin{minipage}{0.45\textwidth}
		\centering
		
		\includegraphics[width= 80mm]{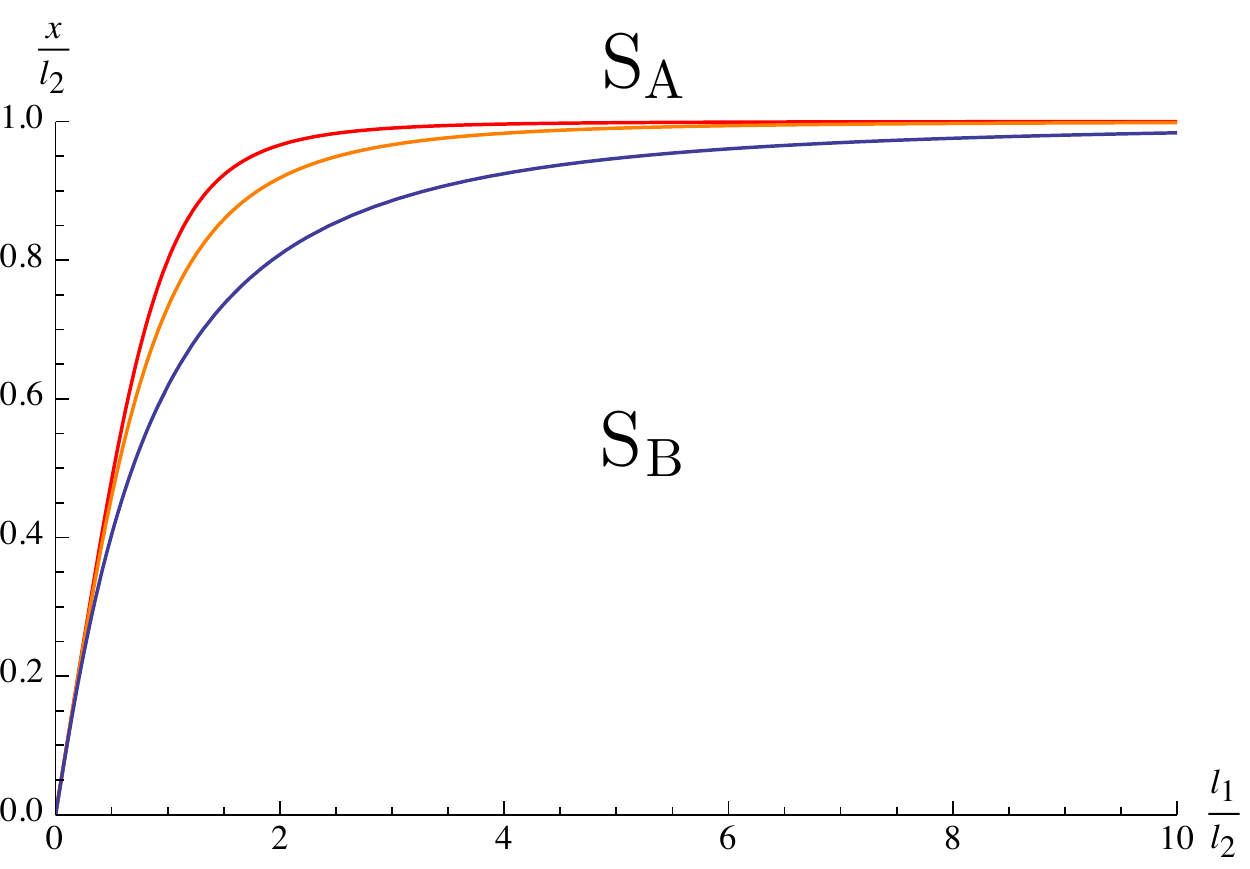}
		\caption[cft2stripsunequalseparation]{ The phase diagram for 2 strips with unequal length. The curves are the transition lines. The curves from top to bottom correspond to dimensions $d=5,4 $ and $3$. The corresponding bulk surfaces are illustrated in Fig~\ref{cft2stripsunequalseparation}. \label{2stripsdifferentlengthdiagram}}
	\end{minipage}
\end{figure}

Now let's consider the case of 2 strips with different lengths $l_1$ and $l_2$, and a separation $x$ (so now there is an additional scale). Again we have just two bulk surfaces $S_A$ and $S_B$ as shown in Fig.~\ref{cft2stripsunequalseparation}.
The areas of the bulk surfaces are (see Eqs.~\ref{eq:qq2}-\ref{eq:qq3}):
\begin{align} 
S_{A}(x,l)&= S_1(l_1)+ S_1(l_2) = - \frac{1}{l_1^{d-2}}- \frac{1}{l_2^{d-2}}\\
S_{B}(x,l)&= S_1(x)+S_1\big(x+l_1+l_2\big) = -  \frac{1}{x^{d-2}} - \frac{1}{(x+l_1+l_2)^{d-2}} 
\end{align}


The phase diagram is shown in Fig~\ref{2stripsdifferentlengthdiagram}. The three transition lines correspond to different values of $d$. Note that when $l_1/l_2 \to \infty$ the phase transition will occur at $x=l_2$ (where $l_2$ is the length of the smaller strip). The transition of eq.~\ref{eq:qq1}  (two strips of equal length) corresponds to the point in the plot where $\frac{l_1}{l_2}=1$. 

\begin{figure}
	\centering
	\begin{minipage}{0.45\textwidth}
		\centering
		
		\includegraphics[width= 80mm,height=45mm]{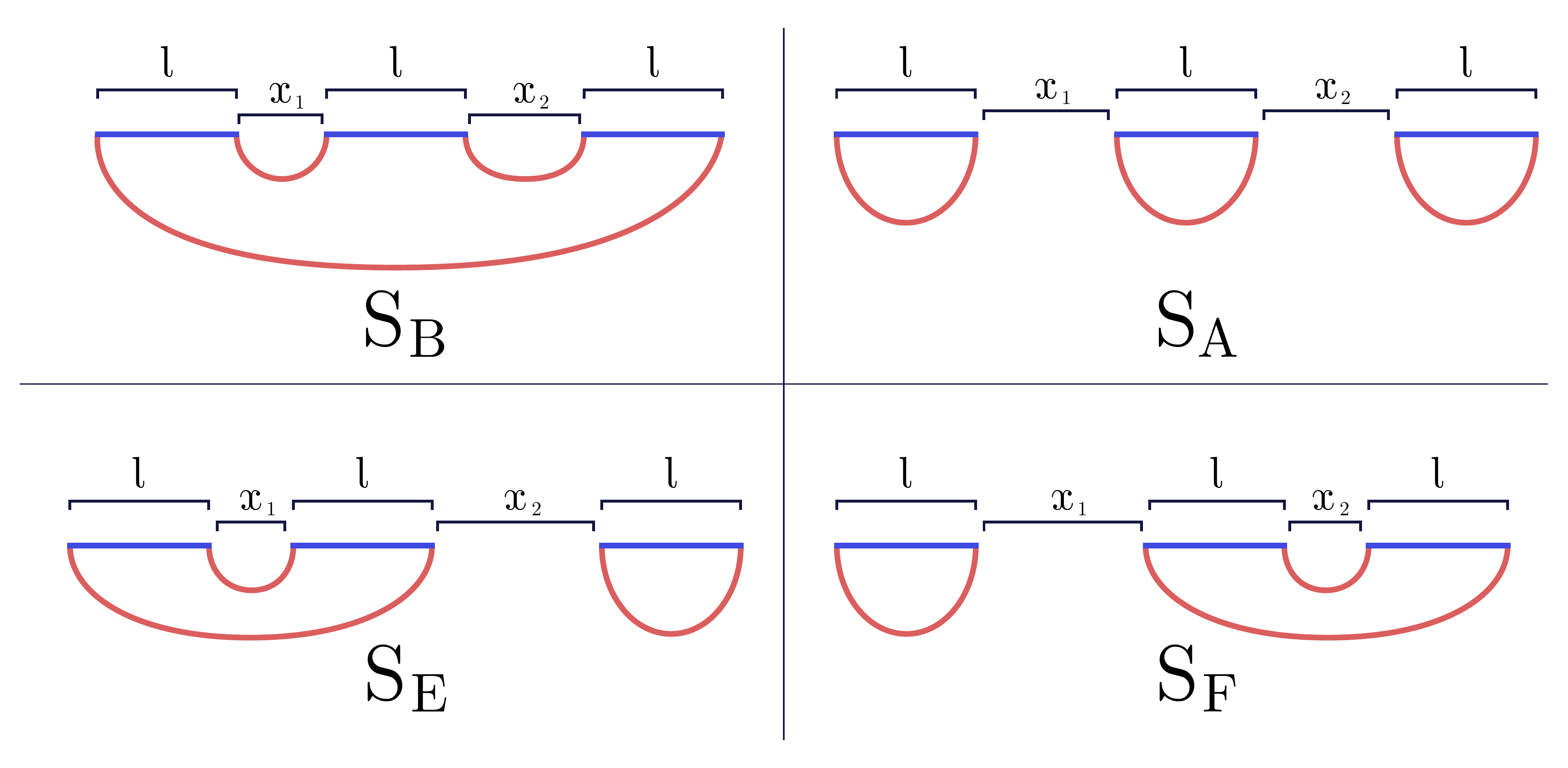}
		\caption[sec:book2]{Bulk surfaces for 3 strips of equal lengths $l$ and unequal separations $x_1$ and $x_2$. $S_E$ and $S_F$ can only be minimal when $x_1\neq x_2$, as proved in section~\ref{sec:theorems4}  . }\label{threestripsdifferentseplength}
	\end{minipage}\hfill
	\begin{minipage}{0.45\textwidth}
		\includegraphics[width= 70mm]{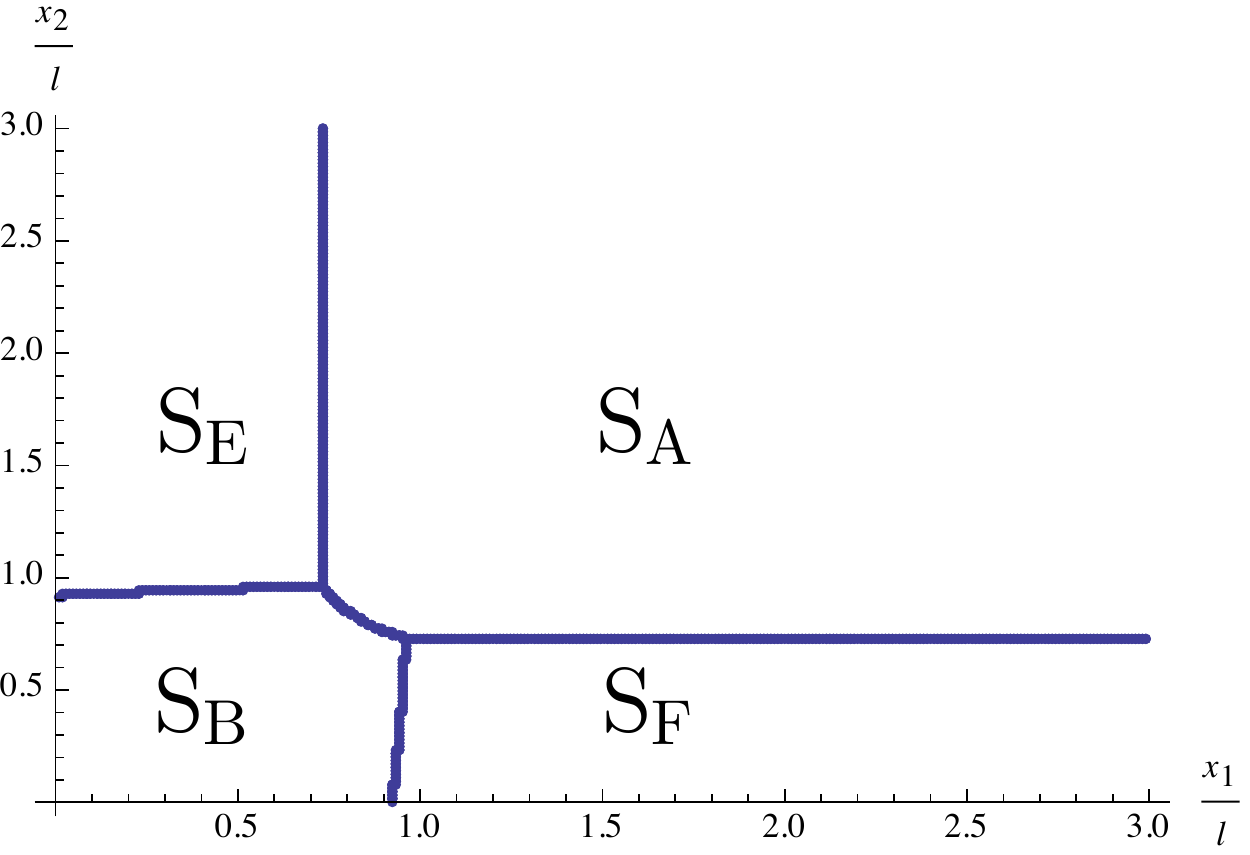}
		\caption[threestripsdifferentseplength]{3 strips of equal lengths $l$ and unequal separations $x_1$ and $x_2$ in $d=4$. The regions correspond to the bulk surfaces of Fig~\ref{threestripsdifferentseplength}. Notice that the plot is symmetrical as it should be. \label{cft3stripsunequalseparation}}
	\end{minipage}
\end{figure}

\subsection{CFT with 3 equal length strips and unequal separations}
\label{sec:book3}
Consider 3 strips of equal length $l$, but with unequal arbitrary separations between them $x_1$ and $x_2$, see also \cite{Allais:2011ys,Balasubramanian:2011at,Alishahiha:2014jxa,Hayden:2011ag}. We now have 2 additional bulk minimal surfaces (denoted $S_E$ and $S_F$) in which one of the strips is ``disjoint" from the other two, see Fig~\ref{threestripsdifferentseplength}. The areas of the 4 bulk surfaces are:
\begin{align} 
S_{A}(x_1,x_2,l)&= 3S_1(l) = - \frac{3}{l^{d-2}}\nonumber\\
S_{B}(x_1,x_2,l)&= S_1(x_1)+S_1(x_2)+S_1(x_1+x_2+3l) = - \frac{1}{x_1^{d-2}} - \frac{1}{x_2^{d-2}} - \frac{1}{(x_1+x_2+3l)^{d-2}} \nonumber\\
S_{E}(x_1,x_2,l)&= S_1(l)+ S_1(x_1+2l) = - \frac{1}{l^{d-2}}- \frac{1}{x_1^{d-2}}- \frac{1}{(x_1+2l)^{d-2}} \\
S_{F}(x_1,x_2,l)&= S_1(l)+ S_1(x_2+2l) = - \frac{1}{l^{d-2}}- \frac{1}{x_2^{d-2}} - \frac{1}{(x_2+2l)^{d-2}} \nonumber
\end{align}

 The phase diagram with its 4 phases is shown in  Fig~\ref{cft3stripsunequalseparation}. To explain the diagram we can start by looking at two small separation lengths ($x_1$ and $x_2$) which is the region near the origin of the diagram. Obviously the minimal surface is the ``joint" one $S_B$. Moving to the right (enlarging $x_1$), one of the strips will disconnect and there will be a transition to $S_F$. Then moving up in the diagram there will be a transition to $S_A$.
There is also a direct transition between $S_A$ and $S_B$ (as for the equal strip case).
Note that the transition line between $S_F$ and $S_A$ is parallel to the $x$-axis,  this is because once the transition to $S_F $ occurred the transition to $S_A$ does not depend on $x_1$. The phase diagram is symmetric under $x_1 \leftrightarrow x_2$ as expected. 

It is not hard to generalize this (but harder to draw) for $m$ strips of equal length and unequal separation. For separations very small compared to $l$: $x_1 \ldots x_{m-1} \ll l$, $S_B$ will be the absolute minimum. For $x_1 \ldots x_{m-2} \ll l$ and $x_{m-1}\gg l$, one strip will be separated form the rest, and so on.

\section{Phases of HEE in confining backgrounds with multiple strips}
 \label{sec:book5}
 So far we have been dealing with bulk theories which are dual to CFTs.  We will now explore backgrounds dual to confining theories \cite{Witten:1998zw}. We first review the 1-strip case, and then we move along to $m$ strips. 

\subsection{Confining background and 1 strip}

In this section we review the 1-strip case, and follow \cite{Klebanov:2007ws}. For more details see Appendix~\ref{app:confinement} and  \cite{Klebanov:2007ws,Nishioka:2006gr,Kol:2014nqa,Nishioka:2009un,Pakman:2008ui,Faraggi:2007fu}.

Consider a bulk metric with the following general form:
 \begin{eqnarray}
 \label{eq:plk}
 ds^2= \alpha_x(U)\big[\beta(U)dU^2+dx^\mu dx_\mu\big]+\alpha_t(U) dt^2+ g^{ij}dy_idy_j
 \end{eqnarray}
 Where $\alpha_x(U)$, $\alpha_t(U)$ and $\beta(U)$ are functions of the holographic direction $U$, $x_\mu$ are the boundary directions ($\mu$ = $1\dots d$) and $y_i$ are internal directions ($i,j=d+2, \ldots , 10$).

 The entanglement entropy is obtained by minimizing the area of the co-dimension 2 bulk surface:
 \begin{eqnarray}
 \label{eq:qq5}
 S=  \frac{1}{4G_N}\int d^{d-1}x e^{-2\phi} \sqrt{\det g^{(ind.)}_{\mu \nu}}\ \ \ \  .
 \end{eqnarray}
Where $\phi$ is the dilaton, and $g_{\mu \nu}^{(ind.)}$  is the induced metric in the string frame.
 
 Considering a strip of length $l$, we plug the metric Eq.~\ref{eq:plk} into Eq.~\ref{eq:qq5} and get: 
 \begin{eqnarray}
  \label{eq:qq8}
 S=  \frac{\tilde{L}^{d-2}}{4G_N}\int_{-l/2}^{l/2} dx \sqrt{H(U)}\sqrt{1+\beta(\partial_x U)^2}\ \ \ \  .
 \end{eqnarray}

where we defined:
 \begin{eqnarray}
 H(U)\equiv e^{-4\phi}V^2_{int}\alpha_x^{d-1}(U) \ \ \  .
 \end{eqnarray}

\begin{figure}
 	\centering
 	\begin{minipage}{0.45\textwidth}
 		\centering
 		\includegraphics[width= 80mm]{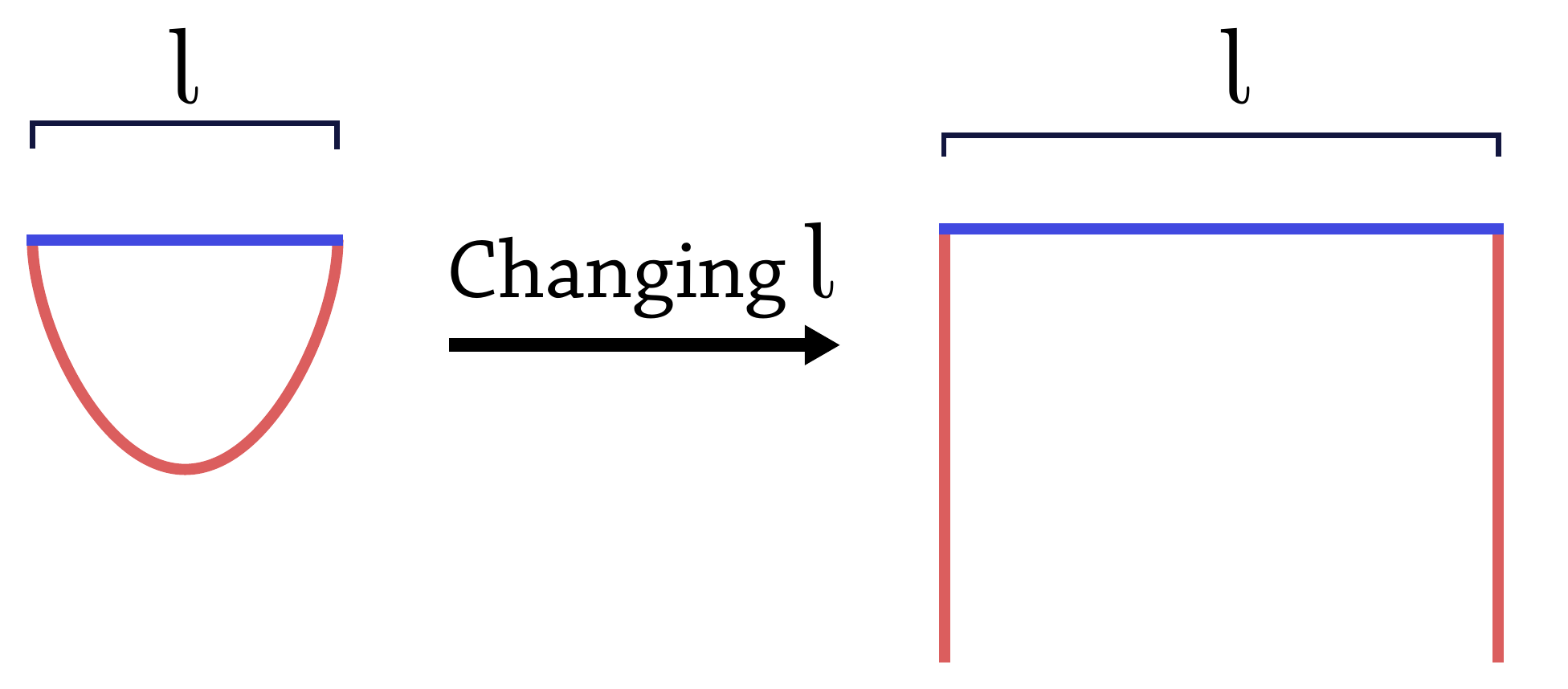}
 		\caption{Illustration of the two minimal surfaces in a confining theory for 1 strip. A phase transition between a ``connected" and ``disconnected" bulk surface. The transition occurs at $l=l_{crit}$.\label{confiningonestrip}}
 	\end{minipage}\hfill
 	\begin{minipage}{0.45\textwidth}
 	\centering
 	\includegraphics[width= 90mm]{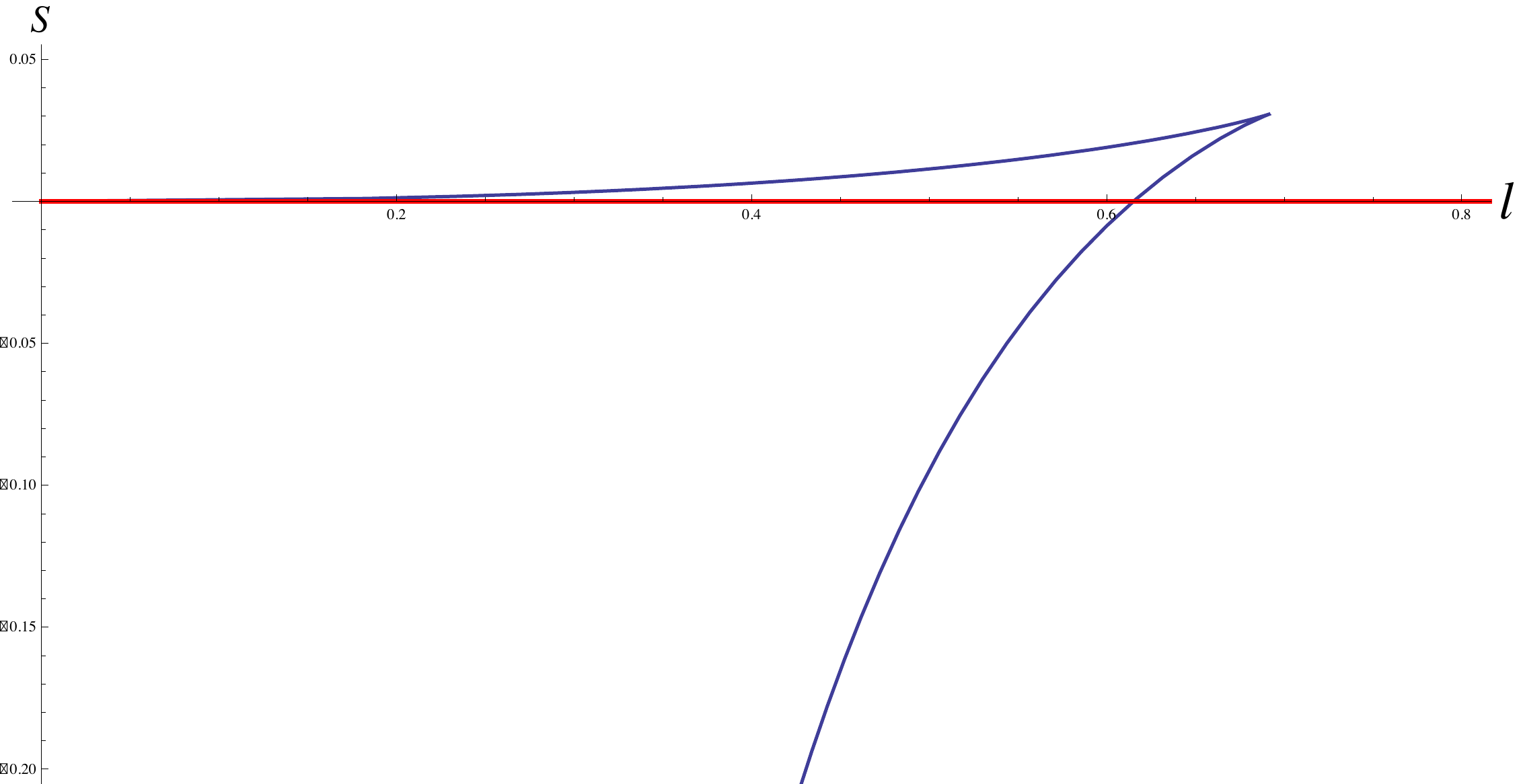}
 	\caption[confiningonestrip]{Showing $S(l)$ for 1 strip in a confining background: $AdS_5$ compactified on a circle. The blue curve is $S^{(conn)}$ and the red curve is $S^{(disconn)}$. The phase transition occurs at $l=l_{crit}\approx 0.61$. An illustration of the bulk surfaces is given in Fig~\ref{confiningonestrip} \label{conf2}}
 	\end{minipage}
 \end{figure}

Confining backgrounds are characterized by a value $U=U_0$ at which the bulk space ends. In the cases that we consider, there will be a circle that shrinks to zero at $U_0$.
There will be 2 competing bulk minimal surfaces, denoted as the ``connected" and ``disconnected" surfaces, see Fig.~\ref{confiningonestrip}. The HEE will correspond to the area of the absolute minimal surface.
The ``disconnected" surface is the surface that goes straight down to the tip of the cigar at $U_0$.
We can obtain the equations of motion from Eq.~\ref{eq:qq8}, and then plug them back into Eq.~\ref{eq:qq8}. We get the area of the ``connected" and ``disconnected" surfaces:
 \begin{eqnarray}
 \label{eq:ry345}
 S^{(conn)}= \frac{\tilde{L}^{d-2}}{2G_N}\int_{U^*}^{U_\infty}\frac{dU \sqrt{\beta(U)}H(U)}{\sqrt{H(U)-H(U^*)}}
 \end{eqnarray}
 
 \begin{eqnarray}
 \label{eq:ry3411}
 S^{(disconn)}= \frac{\tilde{L}^{d-2}}{2G_N}\int_{U_0}^{U_\infty} dU \sqrt{\beta(U)H(U)}
 \end{eqnarray}
 
 \begin{eqnarray}
 \label{eq:ry543}
 l(U^*)= 2 \sqrt{H(U^*)} \int_{U^*}^\infty \frac{dU \sqrt{\beta (U)}}{\sqrt{H(U)-H(U^*)}}
 \end{eqnarray}

$U=U^*$ is where the surface ends, and $U_0$ is the minimal point where the contractible cycle shrinks. Importantly, the area of the disconnected surface is independent of $l$, which enables us to set this constant to 0 (we care only about the differences between the ``connected" and ``disconnected" surfaces).

To illustrate the phase transition, consider the background of $AdS_5 \times S^5$ (D3-branes) compactified on a circle. The metric is:

\begin{align}
ds^2_{10}&=\left(\frac{U}{R}\right)^2\left[\left(\frac{R}{U}\right)^4\frac{dU^2}{f(U)}+dx^\mu dx_\mu\right]+R^2d\Omega^2_5+\left(\frac{U}{R}\right)^2f(U)(dx^3)^2
\end{align} 
Where $f(U)=1-\left(\frac{U_0}{U}\right)^4$\ ,\ $R^4=4\pi\lambda$\ ,\ $U_0^2=\frac{\pi\lambda}{R_3^2}$,  and $\phi=$constant.
This metric is in the form of Eq.~\ref{eq:plk}, with:
\begin{align}
\alpha_x =\alpha_t =\left(\frac{U}{R}\right)^2,\ \ \beta=\left(\frac{R}{U}\right)^4\frac{1}{f(U)},\ \ V_{int}=2\pi^4R_3R^4U\sqrt{f(U)},\ \ H(U)=\left(2\pi^4R_3\right)^2R^4U^6f(U)
\nonumber
\end{align}

We can now plug these functions into Eqs.~\ref{eq:ry345}-\ref{eq:ry543}.
The result is shown in Fig.~\ref{conf2} in which we plot $S(l)$ for this background. Recall the prescription that the HEE corresponds to the absolute minimum solution. There are two ``connected" branches (the blue curves), but the top one is never the minimum solution and therefore not physical. The red curve is the constant area of the ``disconnected" surface. There is a transition between the ``connected" and ``disconnected" solutions at the value $l=l_{crit}\approx 0.61$.  The value of $l_{crit}$ depends on the metric of the confining background, and is proportional to the confinement scale. This transition of the HEE was conjectured to be a consequence of the Hagedorn transition of the dual QFT. \cite{Klebanov:2007ws, Nishioka:2006gr}

\subsection{Confining background with 2 strips of equal length $l$}

\begin{figure}
	\centering
	\includegraphics[width= 140mm]{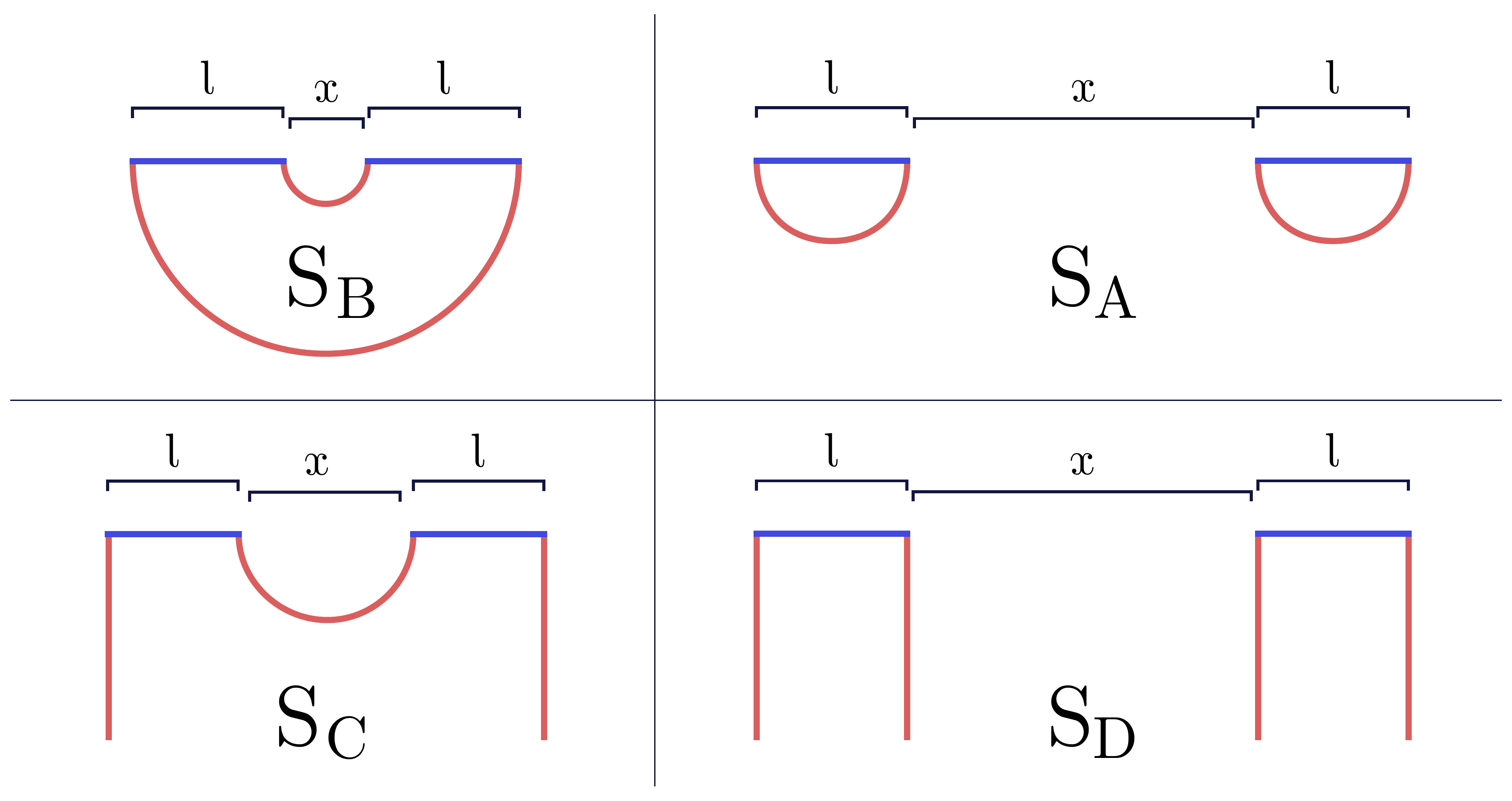}
	\caption{Illustration of the 4 different minimal surfaces for 2 strips of equal lengths $l$ and equal separation $x$ in a confining background. \textbf{Top left:} $S_{B}$: Minimal surface for small $x$ and small $l$. \textbf{Top right:} $S_{A}$: Minimal surface for large $x$ and small $l$. \textbf{Bottom left:} $S_{C}$: Minimal surface for small $x$ and large $l$. \textbf{Bottom right:} $S_{D}$: Minimal surface for large $x$ and large $l$.\label{mu}}
\end{figure}

We will now consider a confining background and 2 strips of equal length $l$ and having a distance $x$ between them. We expect an interplay between the Hagedorn transition mentioned above, and the ``geometric" transitions of section~\ref{sec:mutual1}.
There will be several extremal surfaces, and as usual one has to choose the absolute minimum amongst them for each value of $x$ and $l$. It can be seen in Fig~\ref{mu} that there are 4 different possible minimal surfaces. $S_C$ and $S_D$ are bulk surfaces\footnote{In CFTs these surfaces are never the absolute minimal surfaces.} which contain ``disconnected" pieces as shown Fig~\ref{mu}.

The area of these 4 minimal surfaces can easily be written down:
\begin{align} 
S_{A}(x,l)&= 2S_1(l) & S_{C}(x,l)&= S_1(x)+S_{dis}
\nonumber\\
 S_{B}(x,l)&= S_1(x)+S_1(x+2l)& S_{D}(x,l)&= 2S_{dis}
\end{align}
Where $S_1(l)$ is the area of the ``connected" surface for 1 strip in the confining background obtained from Eq.~\ref{eq:ry345}, and $S_{dis}$ is area of the ``disconnected" surface obtained from Eq.~\ref{eq:ry3411}.

\begin{figure}
\centering
\includegraphics[width= 100mm]{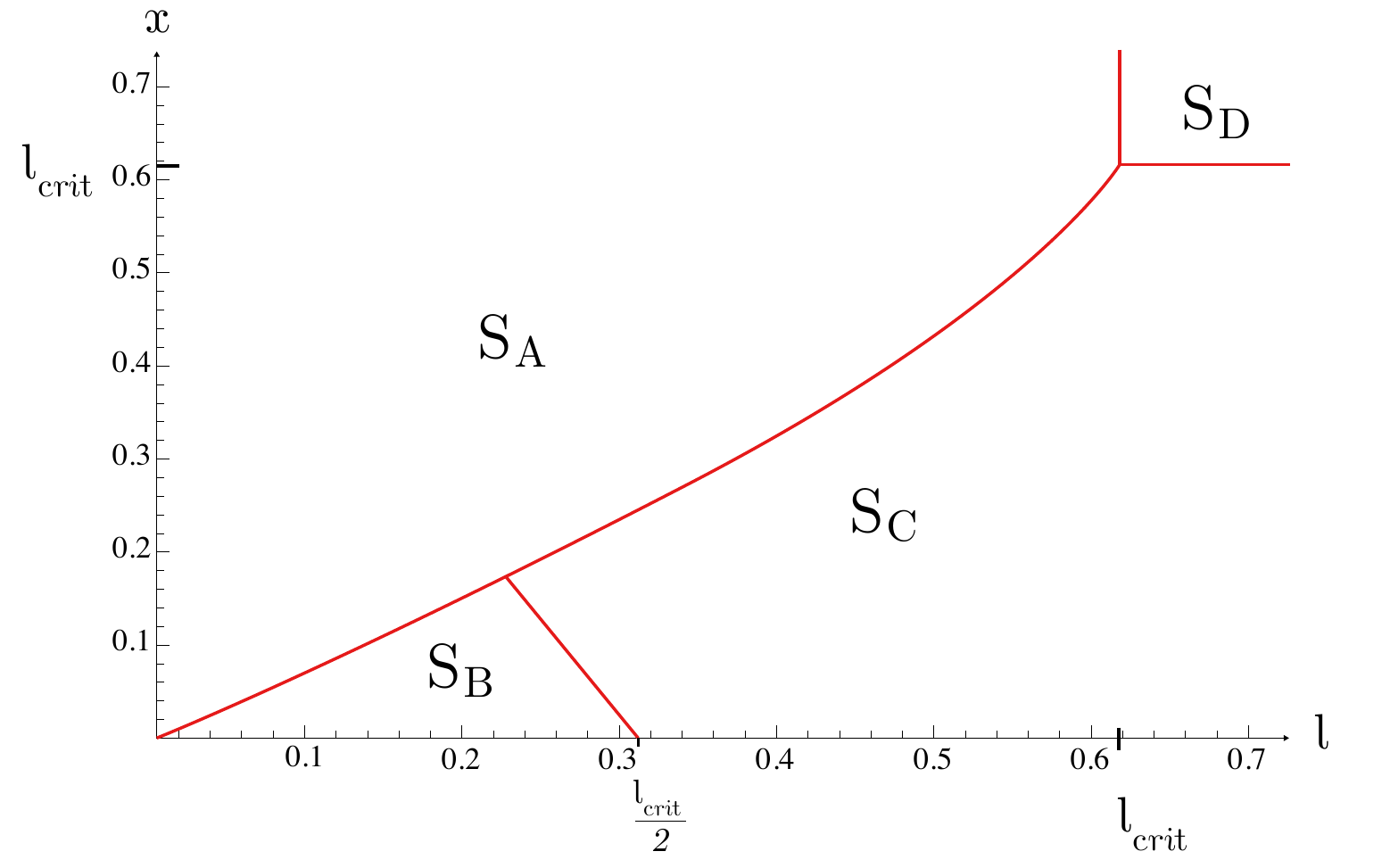}
\caption[mu]{The phase diagram for the background of $AdS_5$ on a circle, and 2 strips of equal lengths $l$ and equal separation $x$. The different phases correspond to the bulk surfaces of Fig~\ref{mu}.\label{mutualphasediagramAdS5}}
\end{figure}

We did the explicit numerical calculation of these functions for the $AdS_5$ (D3 branes) on a circle background, and we show the phase diagram in Fig.~\ref{mutualphasediagramAdS5}. One can intuitively understand the different regions in the plot as follows. The region near the origin (where all lengths are small $x,l \ll l_{crit}$) is the CFT-like region which is similar to that of Fig.~\ref{phasecftK}. It is also clear that in the region $x,l>l_{crit}$, $S_D$ is the minimal surface. Additionally, for small $x$ and $l<l_{crit}$, there is a large region in which $S_C$ is the minimal surface. 

\begin{figure}
	\centering
	\includegraphics[width= 130mm]{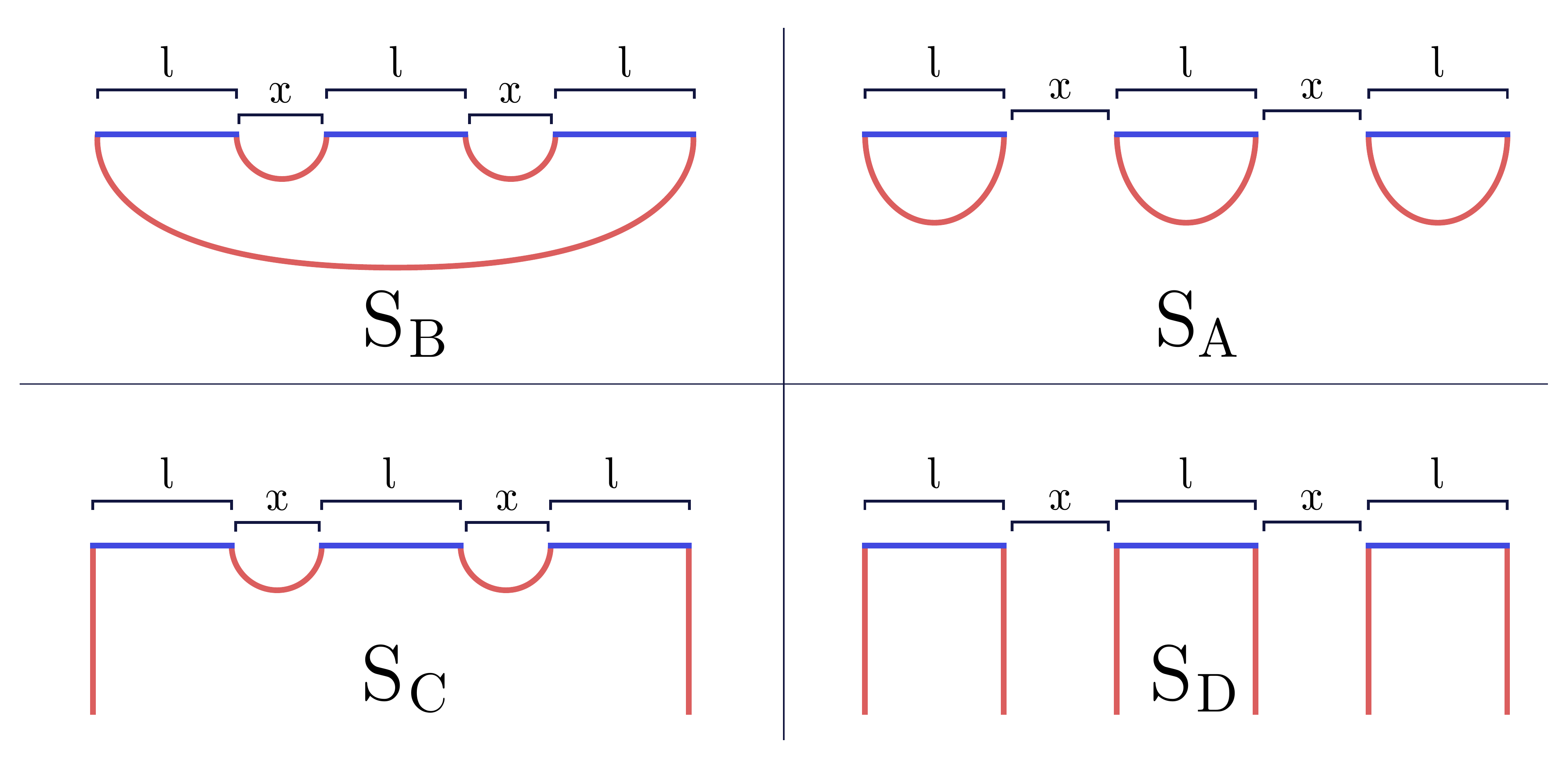}
	\caption{Showing the bulk minimal surfaces for $m$ strips of equal lengths $l$ which have equal separations $x$. Showing 4 different types of minimal surfaces. The plot illustrates the bulk surfaces for $m=3$. \label{3stripsmu3}}
\end{figure}

\subsection{Confining background and $m$ strips}
\label{sec:df3}

\begin{figure}[!h]
	\centering
	\begin{minipage}{0.85\textwidth}
		\centering
		\includegraphics[width= 90mm]{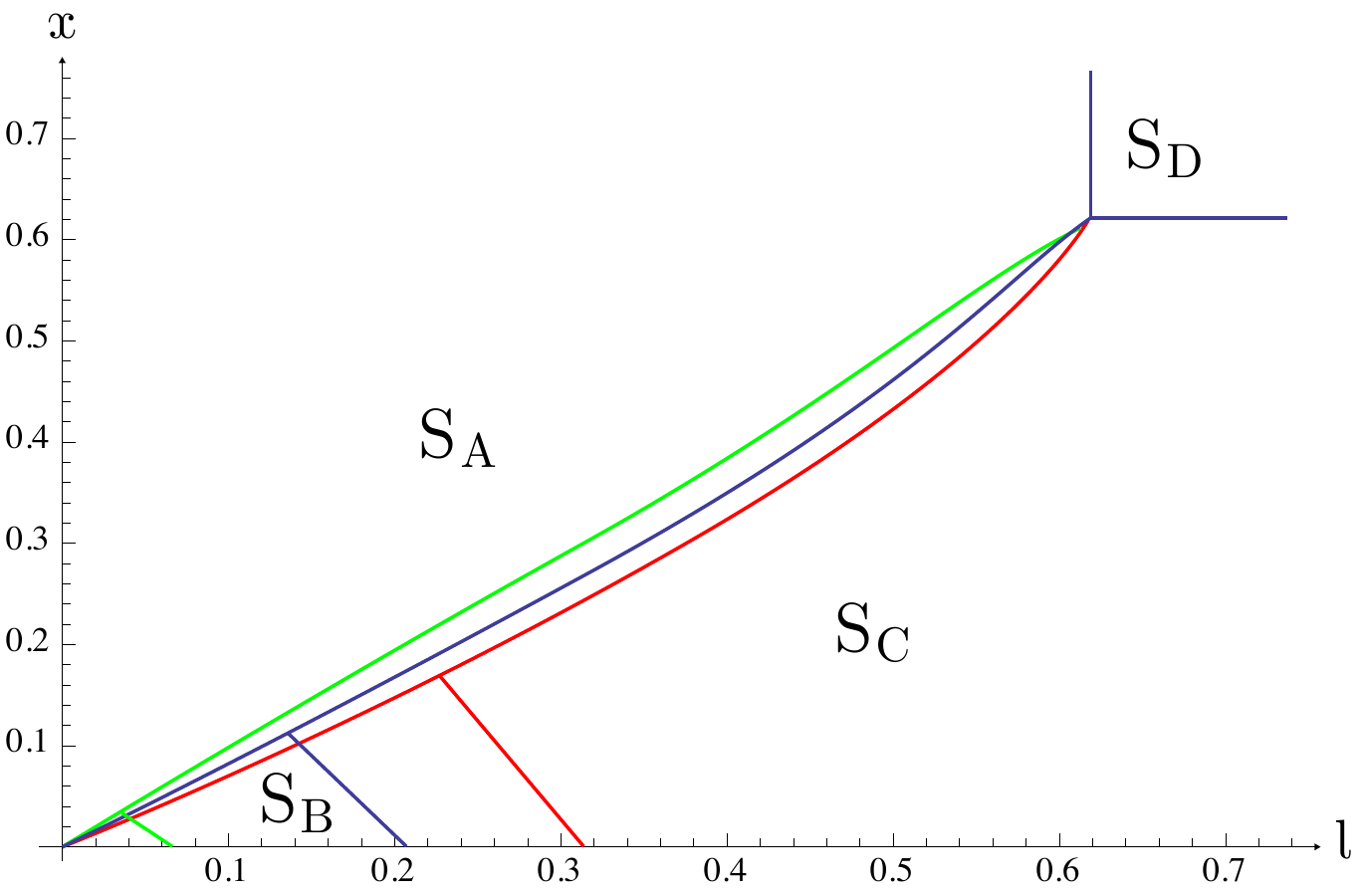}
		\caption[3stripsmu3]{Phase diagram for $AdS_5$ compactified on a circle. Comparing different number of strips $m$. We have $m=2$ (red), $m=3$ (blue), and $m=10$ (green). Illustration of the bulk minimal surfaces is given in Fig~\ref{3stripsmu3}.\label{2stripsVS3stripsVS10strips}}
	\end{minipage}
\end{figure}
Consider the case of $m$ equal length strips with equal spacings between them. In the previous section we saw that for $m=2$ there are 4 bulk minimal surfaces, Fig.~\ref{mu}. For $m>2$, the analogues of these 4 surfaces are shown in Fig.~\ref{3stripsmu3} for the $m=3$ example. In section~\ref{sec:theorems4} it is proved that for $m$ strips with equal lengths and equal separations, $S_A$,  $S_B$,  $S_C$, and $S_D$ are the only bulk minimal surfaces. 
Therefore, the problem of $m$ equally spaced identical strips in a confining background is not more complicated than the 2-strips case. 

The area of these 4 minimal surfaces is, Fig.~\ref{3stripsmu3}:
\begin{align} 
\label{eq:wind2}
S_{A}(x,l)&= mS_1(l) \\\nonumber
S_{C}(x,l)&= (m-1)S_1(x)+S_{dis}\\\nonumber
S_{B}(x,l)&= (m-1)S_1(x)+S_1\big((m-1)x+ml\big)\\
S_{D}(x,l)&= mS_{dis}  \nonumber
\end{align}

We show in Fig.~\ref{2stripsVS3stripsVS10strips} the phase diagrams for $AdS_5$ compactified on a circle and various values of $m$. There is a new feature in this phase diagram. Notice that as we add more strips, the region $S_B$ gets smaller and eventually will disappear for $m\to\infty$.  Relating this to the mutual information, we see that we are left with two ``disjoint" configurations $S_A$ and $S_D$ with zero mutual information, and one ``joint" configuration $S_C$ with mutual information that depends only on $x$, and not on $l$. This is very different from the CFT case where the mutual information is either 0 or depends both on $x$ \underline{and} $l$ (see Eq.~\ref{eq:wind1}). 

The shrinking of the region $S_B$ can easily be seen from Eq.~\ref{eq:wind2}: The transition between $S_B$ and $S_C$ occurs when $(m-1)x+ml > l_{crit}$. For $m \to \infty$ the transition occurs at $x,l \to 0$. Therefore the region $S_B$ shrinks to zero.


\section{``Entanglement plateau-like" transitions and multiple strips}
\label{sec:btz23}

\begin{figure}[!ht]
	\centering
	\begin{minipage}{0.45\textwidth}
		\centering
		\includegraphics[width= 90mm]{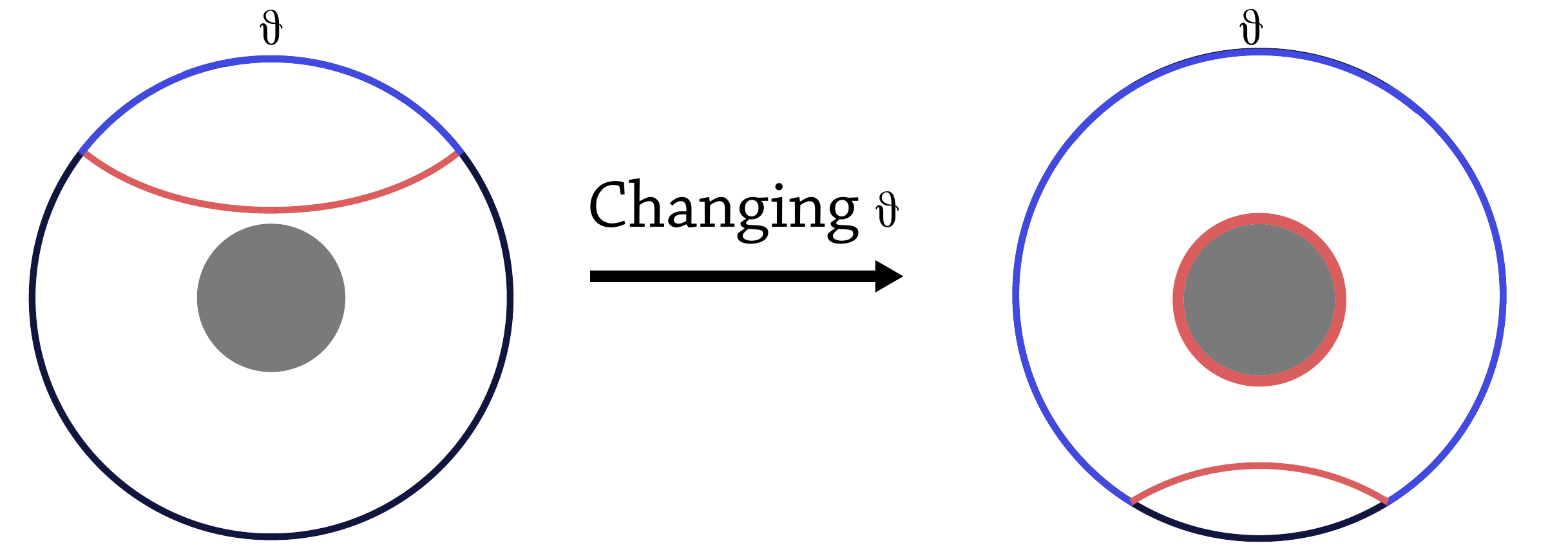}
		\caption{The ``entanglement plateau" transition. An interval of length $\theta$ on the boundary of the global BTZ black hole. The interval is the blue arc, and the minimal surface is the red curve. When $\theta>\theta_c$, the minimal surface contains 2 parts, (one of them wraps the horizon). \label{btzonestrip}}
	\end{minipage}\hfill
	\begin{minipage}{0.45\textwidth}
		\centering
		\includegraphics[width= 80mm]{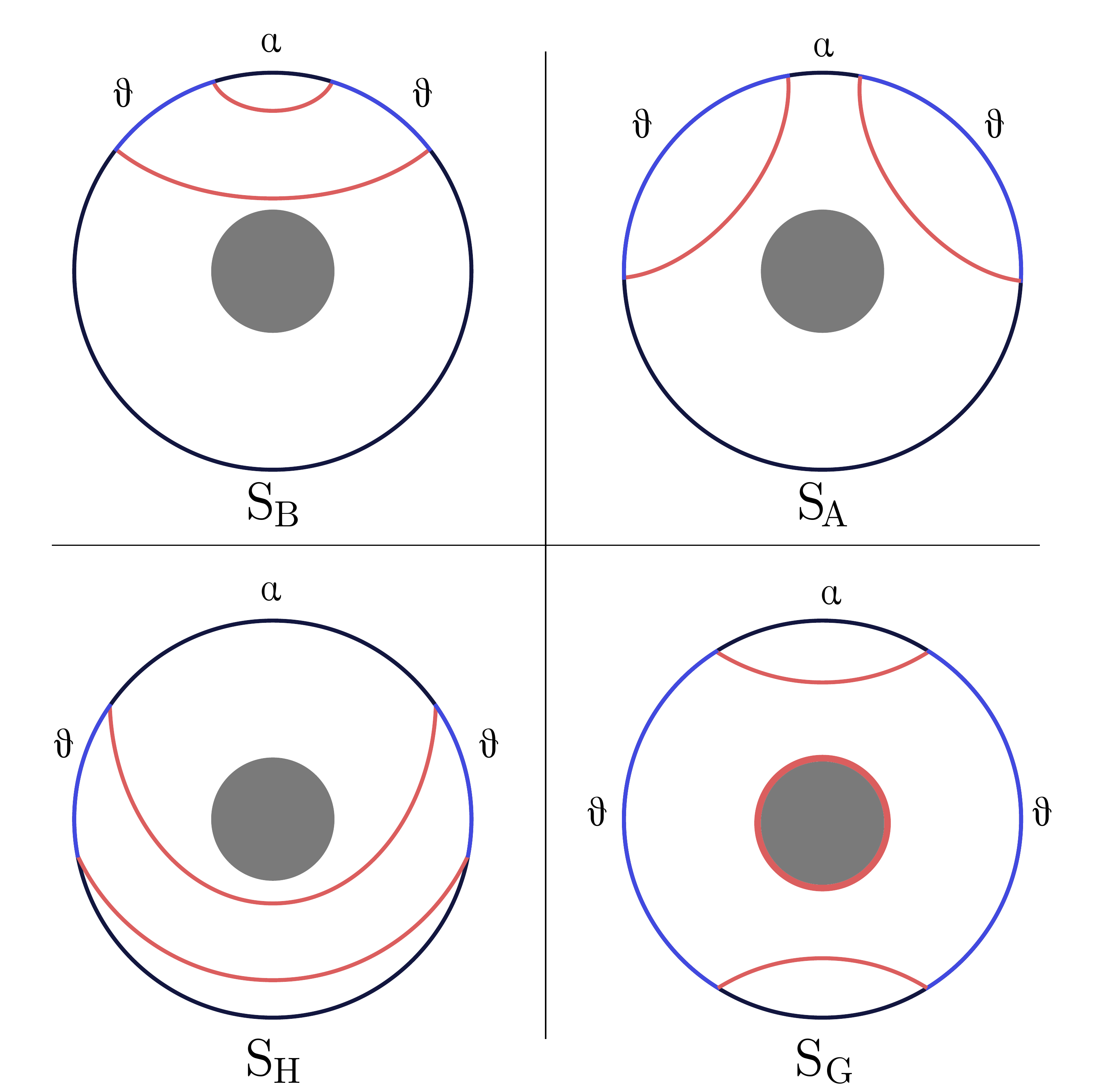}
		\caption{2 equal strips in global BTZ black hole geometry. Illustrating the 4 minimal surfaces $S_A$, $S_B$, $S_G$, $S_H$. $\theta$ is the angle of the strips, and $\alpha$ is the angle separating them. Note that because there are two strips, $\theta$ is smaller than $\pi$.\label{twostripsbtz}}
	\end{minipage}
\end{figure}
\begin{figure}[!ht]
	\begin{minipage}{0.95\textwidth}
		\centering
		\includegraphics[width= 110mm,height=80mm]{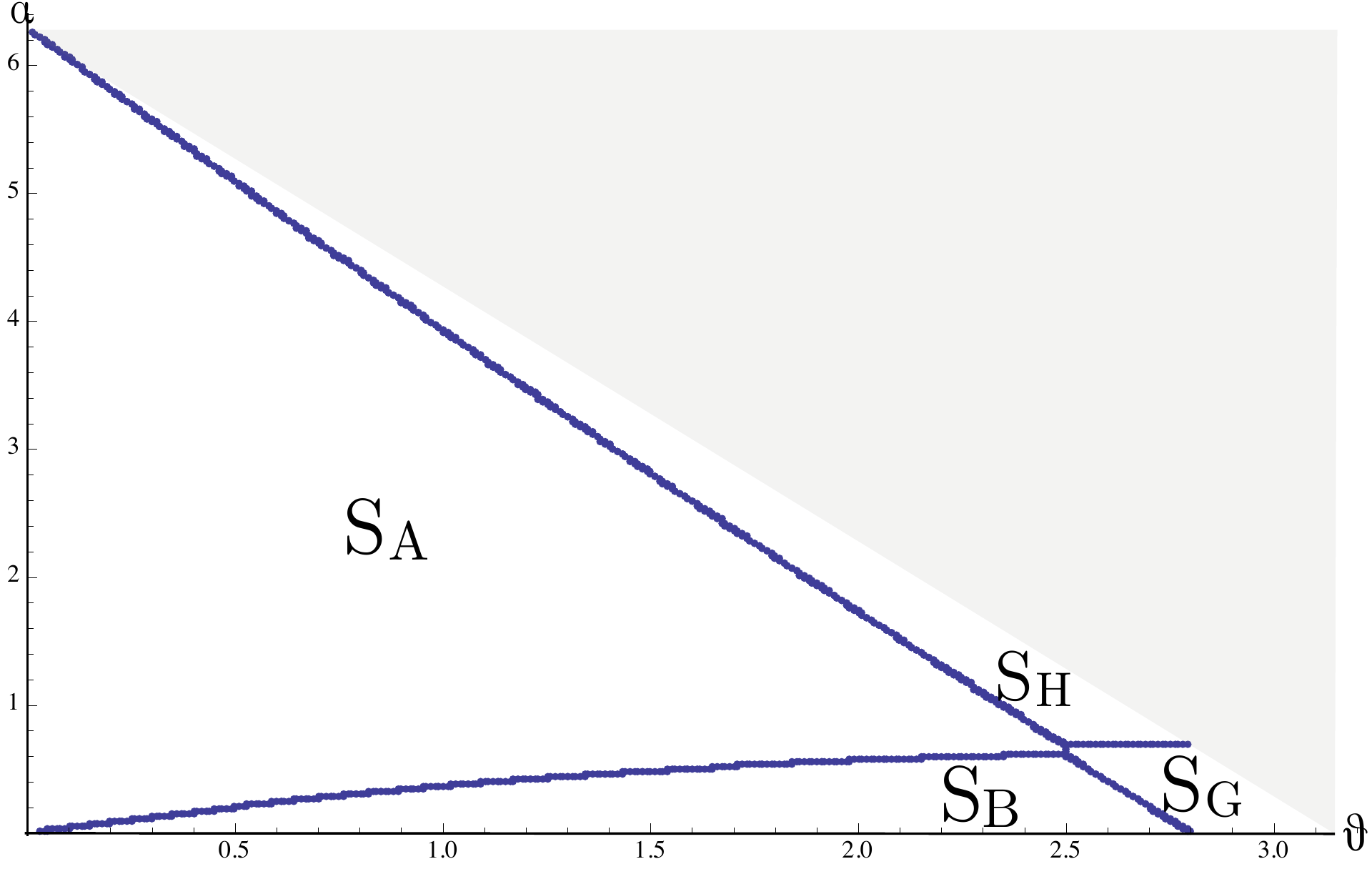}
		\caption[twostripsbtz]{The phase diagram in the $\theta- \alpha$ plane, for the global BTZ black hole and 2 strips. In this plot we used the minimal value $\frac{\pi R_0}{\beta}=\frac{1}{2}$. The four regions correspond to the bulk surfaces illustrated in Fig.~\ref{twostripsbtz}. The gray region is not part of the phase diagram since it is not in the domain $2\theta + \alpha \leq 2 \pi$ . \label{twostripsbtzdiagram}}
		\hfill
	\end{minipage}
	
	\begin{minipage}{0.95\textwidth}
		\centering
		
		\begin{minipage}{0.45\textwidth}
			\centering
			\includegraphics[width= 80mm]{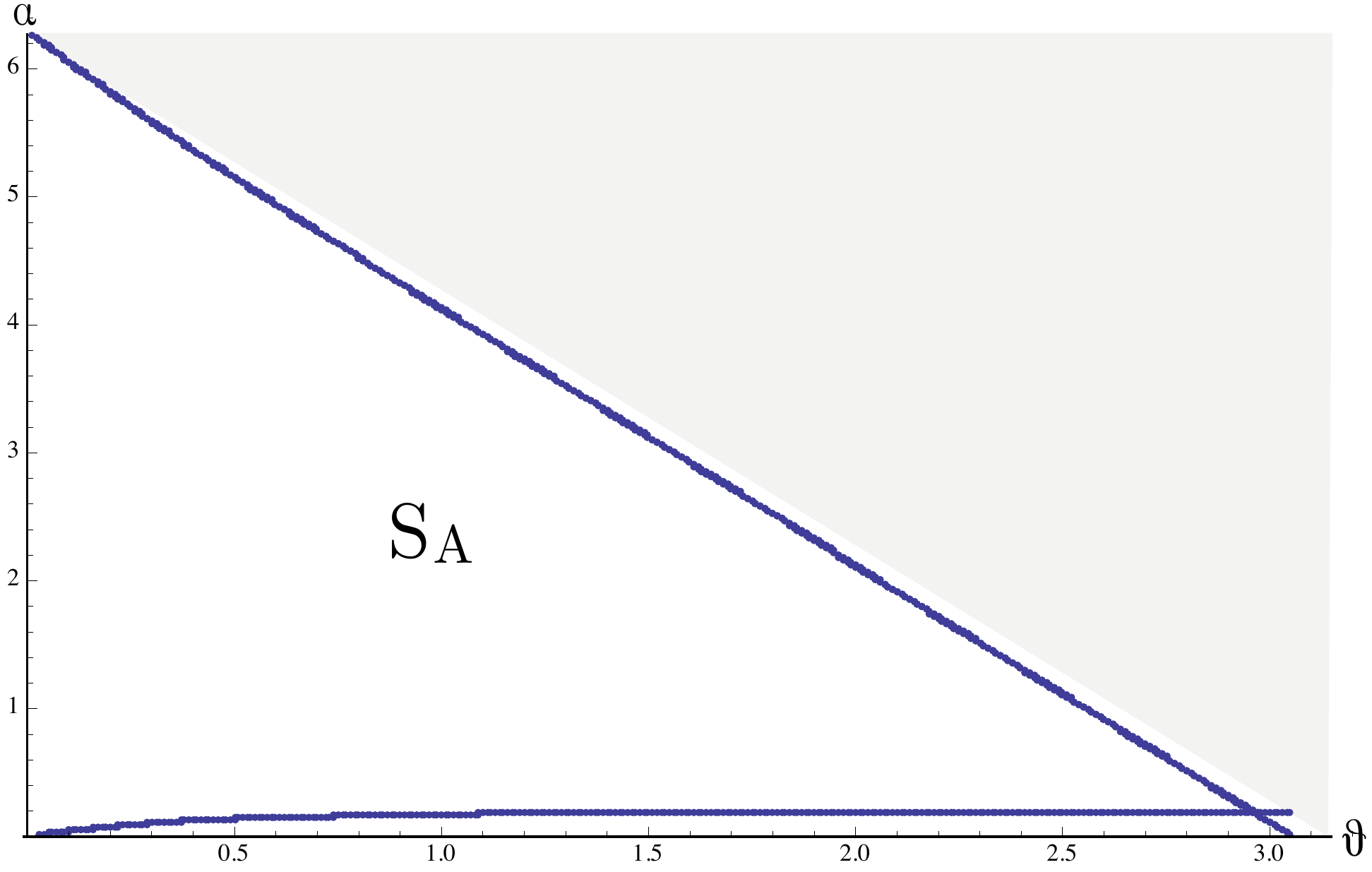}
			
		\end{minipage}
		\hfill
		\begin{minipage}{0.45\textwidth}
			\centering
			\includegraphics[width= 80mm]{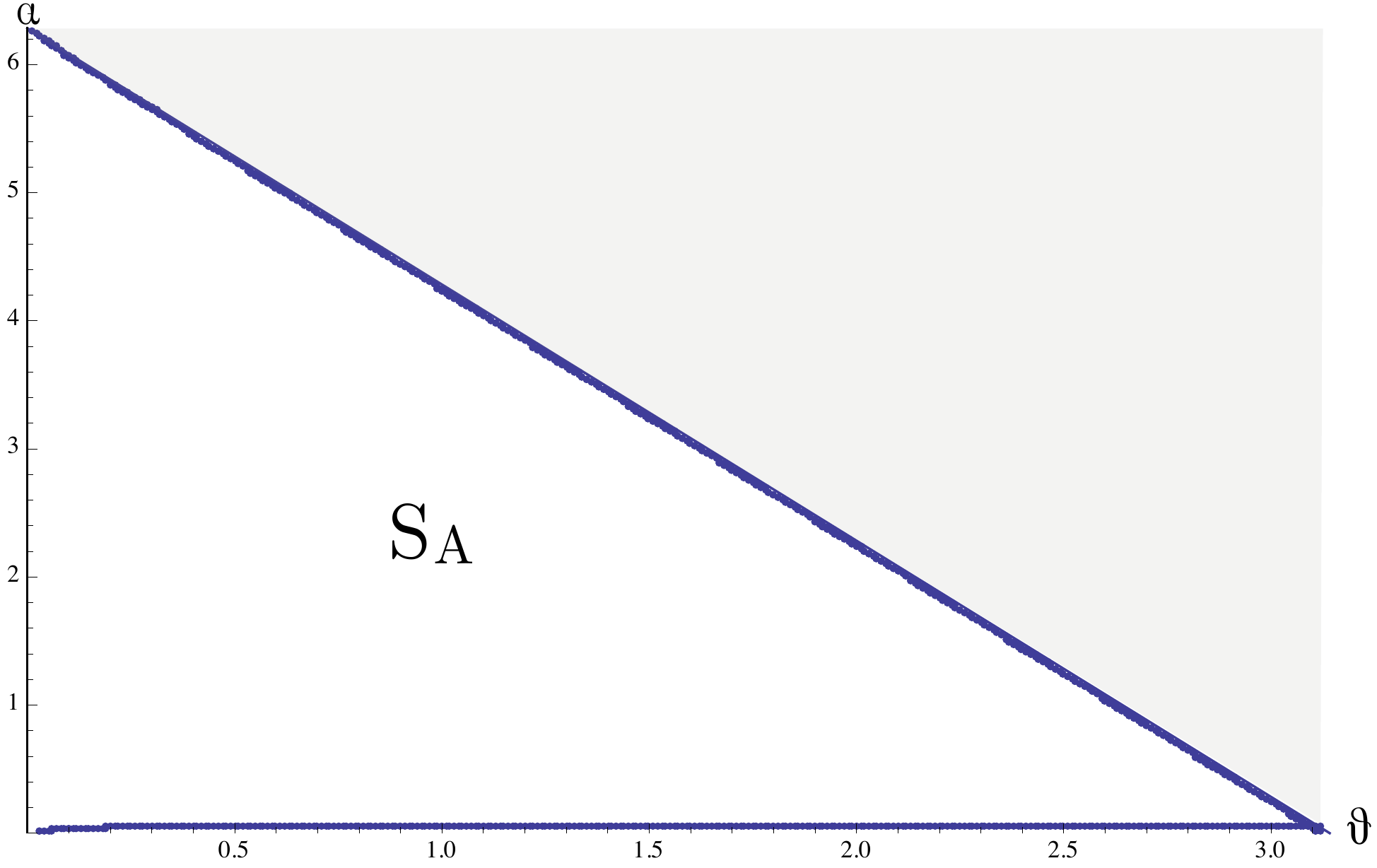}
			
		\end{minipage}
		\caption[twostripsbtzdiagram]{Same as in Fig.~\ref{twostripsbtzdiagram}, but for different sizes of the black hole. It can be seen that enlarging the black hole, the regions $S_G$, $S_H$ and $S_B$ shrink to zero size. \textbf{Left:} $\frac{\pi R_0}{\beta}= 2$. \textbf{Right:} $\frac{\pi R_0}{\beta}= 8$.  \label{twostripsbtzBHc8withlimit}}
	\end{minipage}
\end{figure}

Consider the geometry of a global AdS black hole. For simplicity we will discuss the BTZ black hole in global coordinates:
\begin{eqnarray}
\label{eq:btz54}
ds^2= \frac{r^2-r_+^2}{R_0^2}dt^2 + \frac{L_{AdS}^2dr^2}{r^2-r_+^2} +r^2d \phi^2
\end{eqnarray}
where $L_{AdS}$ is the $AdS$ radius and $R_0$ is the radius of the boundary circle. The inverse temperature is $\beta= 2\pi L_{AdS}R_0/r_+$, and the central charge is $c=12\pi L_{AdS}/l_P$. The dual theory is a 2d CFT with temperature compactified on a circle. There is a Hawking-Page transition: for $T<1/(2\pi R_0)$ the dominant solution is thermal $AdS_3$, and for $T>1/(2\pi R_0)$ the dominant solution is the BTZ black hole Eq.~\ref{eq:btz54}.

Consider an interval region of angle $\theta$ on the boundary circle, Fig.~\ref{btzonestrip}. At a critical value $\theta= \theta_c$ the HEE exhibits a phase transition between two bulk surfaces with different topologies \cite{Hubeny:2013gta,Blanco:2013joa,Johnson:2013dka,Headrick:2007km}. For $\theta>\theta_{c}$ the minimal surface contains 2 ``disjoint" pieces, one of which wraps the horizon of the black hole. The areas of the two minimal surfaces are:
\begin{align}
S_1(\theta)&= \frac{c}{3}\log \Big[\frac{\beta}{\pi \epsilon} \sinh \Big(\frac{\pi R_0 \theta}{\beta} \Big) \Big]\ \ \ , \ \ \ \text{for}\  \theta<\theta_c
\nonumber\\
\tilde{S}_1(\theta)&=S_1(2\pi-\theta)+S_{\text{Horizon}}=  \frac{c}{3}\log \Big[\frac{\beta}{\pi \epsilon} \sinh \Big(\frac{\pi R_0 (2\pi - \theta)}{\beta} \Big) \Big] + \frac{c}{3}\frac{2\pi^2 R_0}{\beta} \ \ \ , \ \ \ \text{for}\  \theta > \theta_c
\end{align}

This transition between $S_1(\theta)$ and $\tilde{S}_1(\theta)$  was called ``Entanglement plateau" in \cite{Hubeny:2013gta}, and it has some similarity to the transition occurring in confining backgrounds, section~\ref{sec:book5}. We note that unlike the case of higher dimensional black holes, the two configurations $S_1(\theta)$ and $\tilde{S}_1(\theta)$  are solutions to the equations of motion for all $\theta$, see \cite{Hubeny:2013gta}.

Now consider 2 strips on the boundary circle, of equal opening angles $\theta$ and a separation $\alpha$. There will be 4 competing bulk surfaces as illustrated in Fig~\ref{twostripsbtz}. 
The areas of these surfaces are:
\begin{align}
\label{eq:windy8}
S_A(\theta, \alpha)&= 2S_1(\theta)
\nonumber\\
S_B(\theta, \alpha)&= S_1(\alpha)+ S_1(2\theta+\alpha)
\nonumber\\
S_G(\theta, \alpha)&= S_1(\alpha)+ S_1(2\pi-\alpha-2\theta)+S_{\text{Horizon}}
\nonumber\\
S_H(\theta, \alpha)&= S_1(2\pi -\alpha)+ S_1(2\pi-\alpha-2\theta)
\end{align}

The phase diagram in the $\alpha-\theta$ plane is shown in Fig.~\ref{twostripsbtzdiagram}, for the minimal value $\frac{\pi R_0}{\beta}=\frac{1}{2}$. In Fig.~\ref{twostripsbtzBHc8withlimit} we show the same phase diagram for the different values $\frac{ \pi R_0}{\beta}=2\text{,}\ 8$. We see that the effect of enlarging $\frac{R_0}{\beta}$ (i.e enlarging the black hole radius) is the shrinking of the regions $S_G$, $S_H$, and $S_B$. Already for $\frac{ \pi R_0}{\beta}=2$, the area of these regions is very small compared to $S_A$.

\section{Phases of HEE in other examples}
\label{sec:book9}

In this section we shortly discuss phases in additional interesting backgrounds.

\subsection{Dp-brane background with two strips of equal length $l$}
The first case we consider is the D$p$-brane background \cite{Itzhaki:1998dd}, which is qualitatively similar to the CFT case. One has to distinguish between the two classes of $p<5$ and $p\geq 5$. Here we discuss just the former class, for the latter see \cite{Barbon:2008ut} \cite{Kol:2014nqa}.

For a single strip in a D$p$-brane background \cite{Pakman:2008ui,Faraggi:2007fu,vanNiekerk:2011yi,Pang:2013lpa}, the finite term is:
\begin{eqnarray}
\label{eq:air1}
S_1(l) = -\frac{1}{l^{\frac{4}{5-p}}}
\end{eqnarray}
Like the CFT case Eq.~\ref{eq:book1}, this is a power law but the exponent now is  non-integer.

Now consider $m$-strips. The analysis is similar to the CFT case, and again we have two competing minimal surfaces $S_A$ and $S_B$ as shown in Fig.~\ref{confutK} for the case $m=2$ .
When $S_{A}=S_{B}$ there will be a transition between the two surfaces. This happens when:
\begin{eqnarray}
\frac{1}{(m+(m-1)y)^{\frac{4}{5-p}}} +  \frac{m-1}{y^{\frac{4}{5-p}}}  = m
\end{eqnarray}
The solution to this equation is $y \equiv x/l=\tilde{f}(p,m)$, where $\tilde{f}(p,m)$ is a function of the $p$ and $m$ only. The phase diagram in the $x-l$ plane will consist of the straight line $x =\tilde{f}(p,m)\cdot l$, qualitatively similar to the CFT case Fig.~\ref{phasecftK} .

\subsection{CFT at finite $T$ with $m$ strips}
\label{sec:df2}
Consider adding temperature to a CFT, \cite{Fischler:2012uv,Tonni:2010pv,Alishahiha:2014jxa}. The mutual information for these theories was computed in \cite{Fischler:2012uv}, where phase diagrams were also determined, hence  we will be very concise here. At a given temperature, the phase diagram still consists of a straight line as in Fig.~\ref{phasecftK}. Enlarging the temperature reduces the slope of the line. For a very large temperature, the slope of the line goes to zero. To see this, note that for large temperature the finite part of the 1-strip EE is given by the thermal entropy\cite{Ryu:2006bv}:
\begin{eqnarray}
S_1(l) \propto T^{d-1} l \ \ \ \ \text{for} \ \ \  \ Tl>>1 
\end{eqnarray}
Therefore for $m$ strips we have:
\begin{align}
S_{A}(x,l)&\propto  m T^{d-1} l\\
 S_{B}(x,l)&\propto   (m-1)T^{d-1} x + T^{d-1}(m l + (m-1)x)\nonumber
\end{align}
Now equating $S_{A}=S_{B}$, we get that the transition line is at $ x=0$. Thus for a very large temperature, the slope of the transition line goes to zero.
Therefore for effectively all values of $x$ and $l$, $S_A$  is the dominant configuration, and there are no correlations between the strips.

\subsection{Wilson loops with $m$ strips}
\label{sec:qq01}

In Appendix \ref{sec:A}, we discuss the similarity between holographic entanglement entropy and holographic Wilson loops in. More precisely, the quark-antiquark potential $V(l)$ at finite temperature has qualitatively similar $l$ dependence as $S_1(l)$ in a confining background.
  
We thus expect that $V(l)$ for a finite $T$ background with $m$ strips will have a phase diagram qualitatively similar to that of the EE in a confining background, namely Fig.~\ref{mutualphasediagramAdS5} . So once more we expect 4 different minimal surfaces.

\section{A general perturbative analysis for $m$ strips}
\label{sec:the2}

\subsection{Perturbing the CFT}
\label{sec:book12}

In this section we start with an $m$-strip configuration in a CFT, and ask what happens when we perturb the CFT. We closely follow \cite{Blanco:2013joa}, see also \cite{Bhattacharya:2012mi}. For CFTs with a temperature turned on see \cite{Fischler:2012uv,Tonni:2010pv,Alishahiha:2014jxa}.

Consider a bulk metric in Fefferman-Graham coordinates:
\begin{eqnarray} 
ds^2 = \frac{R_{AdS}^2}{z^2}\Big(dz^2 + (\eta_{\mu \nu} +\delta g_{\mu \nu})dx^\mu d x^\nu \Big)
\end{eqnarray}

where $R_{AdS}$ is the radius of $AdS$, and the perturbation of the metric is:
\begin{eqnarray} 
\delta  g_{\mu \nu} = \frac{2l_P^{d-1}}{dR_{AdS}^{d-1}}z^d \sum_{n=0} z^{2n}T_{\mu \nu}^{(n)}
\end{eqnarray}
For simplicity assume a uniform stress tensor, i.e $T_{\mu \nu}^{(n)}=0$ for $n\geq 1$.

We can calculate the resulting change (at linear order in the perturbation) in the HEE for 1-strip of length $l$:
\begin{eqnarray} 
\label{eq:air223}
\delta  S_1(l) =\ \frac{1}{2} \int d^{d-1} \sigma \sqrt{g_0}g_0^{ab}\delta g_{ab} \propto  \bigg( \frac{d+1}{d-1}T_{00} - T_{xx}  \bigg)  l^2 \equiv  \varepsilon  l^2 
\end{eqnarray}
where $g_0^{ab}$ is the zeroth order induced metric on the bulk surface, and $\delta g_{ab}$ is its perturbation.
It is important for the following analysis that $\delta  S_1(l) $ is quadratic in $l$ and can be positive or negative depending on the sign of $\varepsilon$.

Let us now consider $m$ equal length strips equally separated. The change in $S_A$ and $S_B$ is:

\begin{align}
\delta S_{A}(x,l)&= m\delta S_1(l)   \propto m \varepsilon l^{2}
\nonumber\\ 
\delta S_{B}(x,l)&= (m-1)\delta S_1(x)+\delta S_1\big((m-1)x+ml\big) \propto (m-1) \varepsilon x^{2} +  \varepsilon \big[(m-1)x+ml\big]^{2}
\end{align}

The corresponding change in the mutual information is either zero or:
\begin{eqnarray}
\delta I=\delta S_{A} - \delta S_{B} = -\epsilon m(m-1)(x+l)^2
\end{eqnarray}
So we see that for a positive/negative  $\epsilon$ the change in mutual information  $\delta I$ is always negative/positive.

We also note that for large $m$:
\begin{eqnarray}
\frac{\delta S_{B}}{\delta S_{A}} \sim  m\Big(1+\frac{x}{l} \Big)^{2} \gg 1
\end{eqnarray}

So depending on whether $\varepsilon$ is positive/negative, $\delta S_{B}$ gets a large positive/negative contribution, much larger (in absolute value) then $\delta S_A$.

Imagine that we start in a CFT with many strips and the bulk minimal surface is $S_B$. Now we perturb the CFT with a ``positive perturbation" $\varepsilon >0$ (such as for a small strip in a background with temperature). Then $\delta S_B$ gets a large positive contribution $\delta S_{B}\sim \varepsilon m^2(x+l )^{2}$, which can render $S_A$ the new minimal bulk surface. Thus a ``positive perturbation" $\varepsilon >0$ tends to break the ``joint" surfaces such as $S_B$ into ``disjoint" surfaces such as $S_A$. This is what happened before in section~\ref{sec:df2} in a background with finite temperature, where the parameter space for $S_B$ became smaller and smaller as we increased the temperature. 

On the other hand, if the perturbation is negative $\varepsilon <0$, then $\delta S_B$ gets a large negative contribution $\delta S_{B}\sim \varepsilon m^2(x+l )^{2}$. Such a negative perturbation tends to join together the ``disjoint" surfaces such as $S_A$.\\

We can also consider other types of perturbations other than the stress energy tensor. We will now mention the cases of a scalar operator and a vector operator. \\

Perturbing with a scalar operator (a scalar field in the bulk), we get (see \cite{Blanco:2013joa} ):
\begin{eqnarray} 
\label{eq:windy6}
\delta  S_1 \propto \mathcal{O}^2 \big( AT_{00} - BT_{xx}  \big)l^{2\Delta -d +2}  \equiv  \varepsilon  l^{2\Delta -d +2} 
\end{eqnarray}
Above the unitarity bound $\Delta >d/2-1$, the exponent of $l$ is positive as in Eq.~\ref{eq:air223}.\\

On the other hand, perturbing with a vector operator we get:
\begin{eqnarray} 
\delta  S_1 \propto \big( CJ^2_{0}+DJ^2_{x} + EJ^2  \big)l^{d }  \equiv  \varepsilon  l^{d} 
\end{eqnarray}

where $A$, $B$, $C$, $D$, and $E$ are constants. The exponent of $l$ is again positive.




\subsection{Perturbing the separations and lengths of the strips}
\label{sec:book112}
\begin{figure}
	\centering
	\includegraphics[width=150mm]{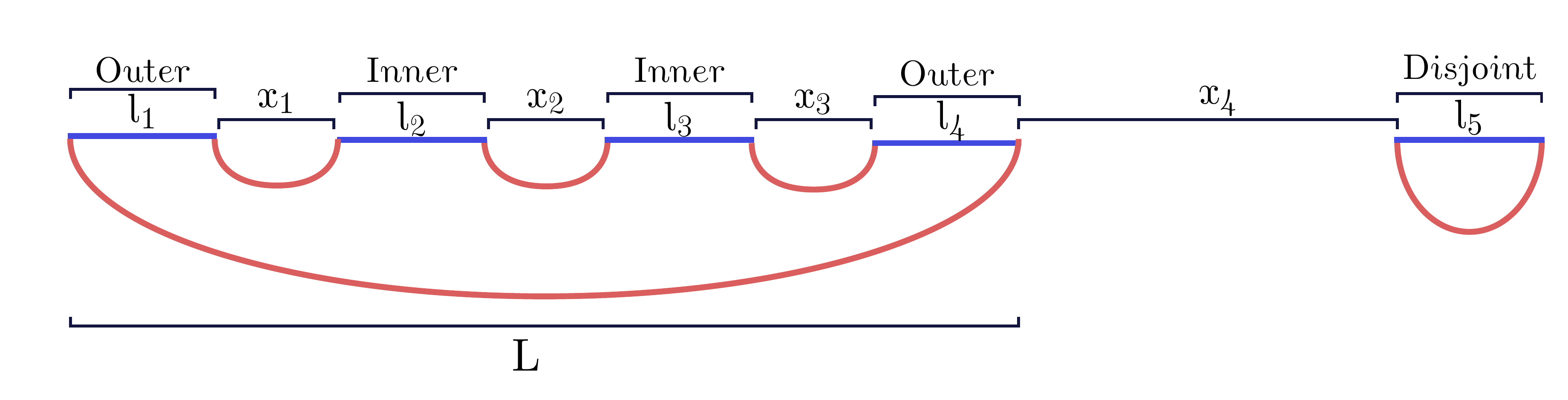}
	\caption{A generic bulk surface illustrating our conventions. A strip such as $l_5$ will be called a ``disjoint" strip. A strip such as $l_1$ will be called an ``outer" strip. A strip such as $l_2$ will be called an ``inner" strip. $L$ is the total length of the ``joint" part of the surface.}
	\label{outerinner}
\end{figure}

In the previous section we saw the effect of perturbing the CFT on the HEE of $m$ strips. In this section we study the effect of slightly changing the length or separation of one of the strips. We assume that these perturbations are small enough as to not cause a phase transition to a different bulk surface.






Consider an arbitrary theory and a generic bulk surface\footnote{We assume here that the bulk surface does not have ``disconnected" parts. However, one can easily redo the analysis of this section for bulk surfaces containing ``disconnected" parts.}, as illustrated in Fig.~\ref{outerinner}. The area $S$ of this surface is:
\begin{eqnarray}
S(x_j,l_j)= S_1(x_1) + S_1(x_2) + S_1(x_3)  +S_1(L) + S_1(l_5)
\end{eqnarray}

We can take derivatives to see how $S$ changes:
\begin{eqnarray}
\label{eq:airt1}
\frac{\partial S}{\partial x_4}= 0
\ \ \ \ \ \ \ , \ \ \ \ \ \ \ \ \ 
\frac{\partial S}{\partial l_5}= \frac{\partial S_1(l_5)}{\partial l_5}
\ \ \ \ \ \ \ , \ \ \ \ \ \ \ \ \ 
\frac{\partial S}{\partial x_1}= \frac{\partial S_1(x_1)}{\partial x_1} + \frac{\partial S_1(L)}{\partial L}
\end{eqnarray}

\begin{eqnarray}
\label{eq:airt2}
\frac{\partial S}{\partial l_1}= \frac{\partial S_1(L)}{\partial L}
\ \ \ \ \ \ \ \ , \ \ \ \ \ \ \ \ 
\frac{\partial^2 S}{\partial x_1 \partial l_1}=  \frac{\partial^2 S_1(L)}{\partial L^2}
\ \ \ \ \ \ \ ,\ \ \ \ \ \ \ \ \  
\frac{\partial^2 S}{\partial x_1 \partial x_2}=  \frac{\partial^2 S_1(L)}{\partial L^2}
\end{eqnarray}

\subsubsection{Changing the separations between the strips}

Now lets see the effect of changing the separations between strips. In the following, we will allow one of the strips to slightly move to the left or to the right (without changing its length). \\

$\bullet$  \ \textbf{A ``disjoint" strip:}

Slightly moving a ``disjoint" strip (see Fig.~\ref{outerinner}) a distance $\Delta x_4$, we get from Eq.~\ref{eq:airt1}:
\begin{eqnarray}
\Delta S =\frac{\partial S}{\partial x_4} \Delta x_4= 0
\end{eqnarray}

$\bullet$  \ \textbf{An ``outer" strip:}

Slightly moving an ``outer" strip (see Fig.~\ref{outerinner}) a distance $\Delta x_1$, we get from Eq.~\ref{eq:airt1}:
\begin{eqnarray}
\Delta S= \frac{\partial S}{\partial x_1}\Delta x_1= \Big( \frac{\partial S_1}{\partial x_1} + \frac{\partial S_1}{\partial L}   \Big) \Delta x_1
\end{eqnarray}
 If $S_1$ is monotonically growing then $\Big( \frac{\partial S_1}{\partial x_1} + \frac{\partial S_1}{\partial L}   \Big)>0$ , hence the sign of $\Delta S$ is the same as that of $\Delta x_1$. Therefore enlarging $x_1$, makes $S$ larger.\\

$\bullet$  \ \textbf{An ``inner" strip:}

Slightly moving an ``inner" strip (see Fig.~\ref{outerinner}) a distance $\Delta x_1$, we get (at linear order):
\begin{eqnarray}
\Delta S^{(1)}= \frac{\partial S_1}{\partial x_1}\Delta x_1 + \frac{\partial S_1}{\partial x_2}\Delta x_2 = \Big( \frac{\partial S_1}{\partial x_1} - \frac{\partial S_1}{\partial x_2}\Big)\Delta x_1
\end{eqnarray}
where we used $\Delta x_1 =-\Delta x_2$ (we keep the total length $L$ fixed).

When $x_1=x_2$, we have $\Delta S^{(1)}=0$,  and $S$ has a maximum at this point. To see this, consider the 2nd order variation:
\begin{eqnarray}
\Delta S^{(2)}= \Big[ \frac{1}{2}\frac{\partial^2 S_1}{\partial x_1^2} + \frac{1}{2}\frac{\partial^2 S_1}{\partial x_2^2}\Big](\Delta x_1)^2
\end{eqnarray}
 
 If $S_1$ is concave then $\frac{\partial^2 S_1}{\partial x_1^2},\ \frac{\partial^2 S_1}{\partial x_2^2}<0$ , hence
\begin{eqnarray}
\Delta S^{(2)} <0
\end{eqnarray}
Therefore, interestingly, $S$ has a maximum when $x_1=x_2$ with respect to slightly moving ``inner" strips.

\subsubsection{Changing the length of the strips}

$\bullet$  \ \textbf{A ``disjoint strip":}

Slightly changing the length of a ``disjoint" strip (see Fig.~\ref{outerinner}) an amount $\Delta l_5$, we get from Eq.~\ref{eq:airt1}:
\begin{eqnarray}
\label{eq:yu1}
\Delta S= \frac{\partial S}{\partial l_5}\Delta l_5 = \frac{\partial S_1(l_5)}{\partial l_5}\Delta l_5
\end{eqnarray}
If $S_1$ is  monotonically growing $\frac{\partial S_1(l_5)}{\partial l_5}>0$, and we enlarge the strip $\Delta l_5>0$, then  $\Delta S>0$.\\

$\bullet$  \ \textbf{An ``outer" strip:}

Slightly changing the length of an ``outer" strip (see Fig.~\ref{outerinner}) an amount $\Delta l_1$, we get from Eq.~\ref{eq:airt2}:
\begin{eqnarray}
\label{eq:yu2}
\Delta S= \frac{\partial S}{\partial l_1}\Delta l_1  = \frac{\partial S_1}{\partial L}\Delta l_1 
\end{eqnarray}
If $S_1$ is  monotonically growing $\frac{\partial S_1}{\partial L}\ > 0$ , and we enlarge the strip $\Delta l_1>0$, then  $\Delta S>0$.\\

$\bullet$  \ \textbf{An ``inner" strip:}

Slightly changing the length of an ``inner" strip (see Fig.~\ref{outerinner}) an amount $\Delta l_2$, we get from the fact that $\Delta l_2 = -\Delta x_2$ (we keep the total length $L$ fixed):
\begin{eqnarray}
\Delta S= \frac{\partial S}{\partial l_2}\Delta l_2  = -\frac{\partial S_1(x_2)}{\partial x_2}\Delta l_2 
\end{eqnarray}
If $S_1$ is  monotonically growing $\frac{\partial S_1(x_2)}{\partial x_2}>0$, and if we enlarge the strip $\Delta l_2>0$,\ then  $\Delta S<0$.
Interestingly, this is opposite behavior compared to a ``disjoint" strip or an ``external" strip, see Eqs.~\ref{eq:yu1} and \ref{eq:yu2}.

\section{Excluding classes of bulk minimal surfaces}
\label{sec:theorems4}

In this section we will exclude certain classes of bulk minimal surfaces. 
For $m$-strips, consider the ``connected" bulk surfaces\footnote{``Disconnected" surfaces are bulk surfaces which have parts that terminate at the end of the bulk space. $S_C$ and $S_D$ are examples of ``disconnected" surfaces. ``Connected" surfaces are bulk surfaces which are not ``disconnected".}. There are $(2m-1)!!$ locally minimal surfaces obtained by all the different pairings of the $2m$ entangling surfaces of the strips. In the following proofs we refer to the plots of the bulk minimal surfaces, but the proofs only use the strong subadditivity property of the EE and do not use the fact that the entangling regions are strips. Therefeore the proofs should work for $m$ identical entangling regions equally separated on a line. \\

\begin{figure}
 	\centering
 	\includegraphics[width= 60mm]{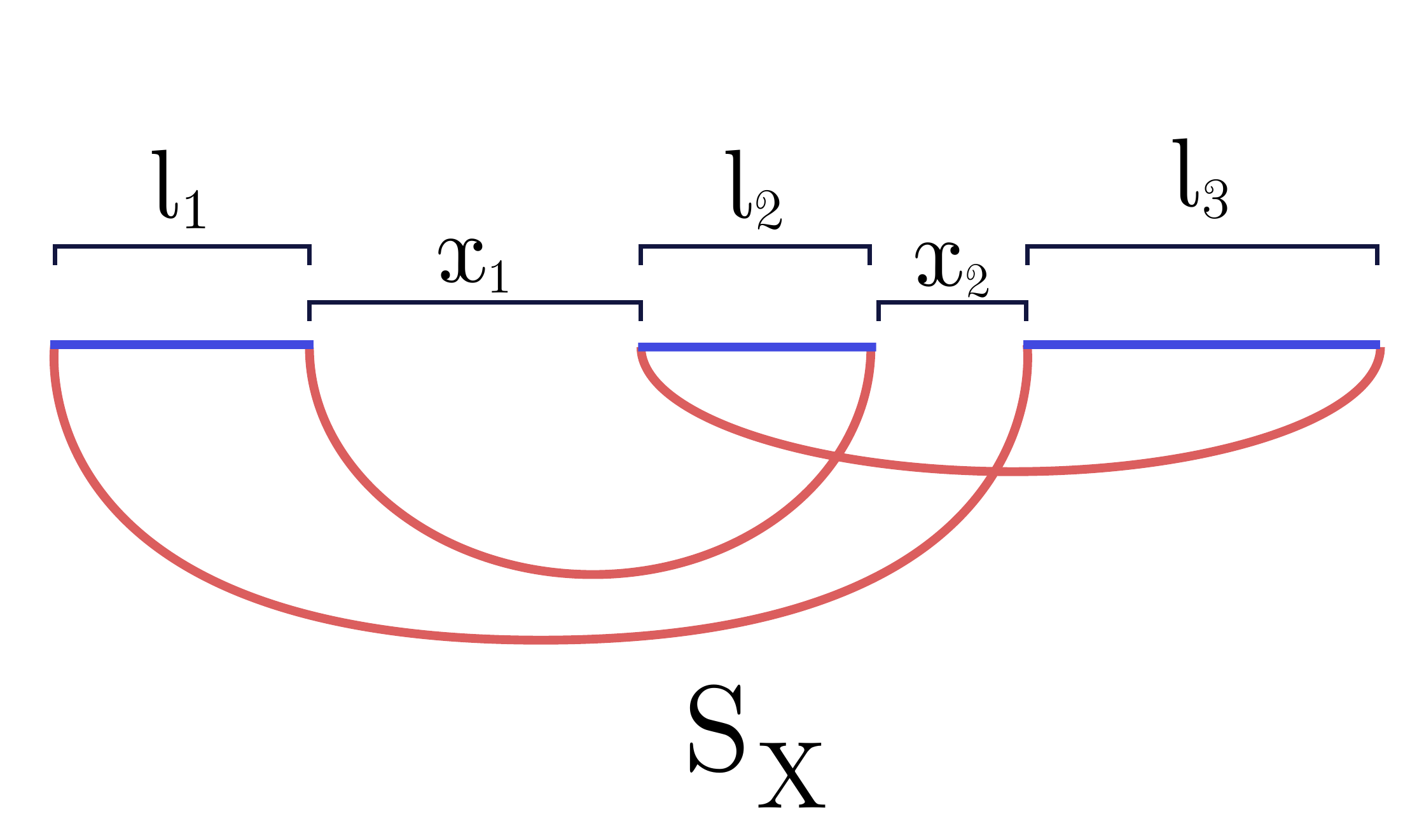}
 	\caption{The class of ``intersecting" bulk minimal surfaces denoted by $S_X$. It is easy to see that these are never the absolute minimal surfaces. \label{sx}}
 \end{figure}

$\bullet$ The class of bulk surfaces (denoted $S_X$) which have intersections Fig.~\ref{sx}, are never the (absolute) minimal surfaces. Proving this is straightforward: For each such intersecting bulk surface it is simple to find a non-intersecting bulk surface with a smaller area. This result has been noted by several authors.\\

\begin{figure}
	\centering
	\includegraphics[width= 120mm]{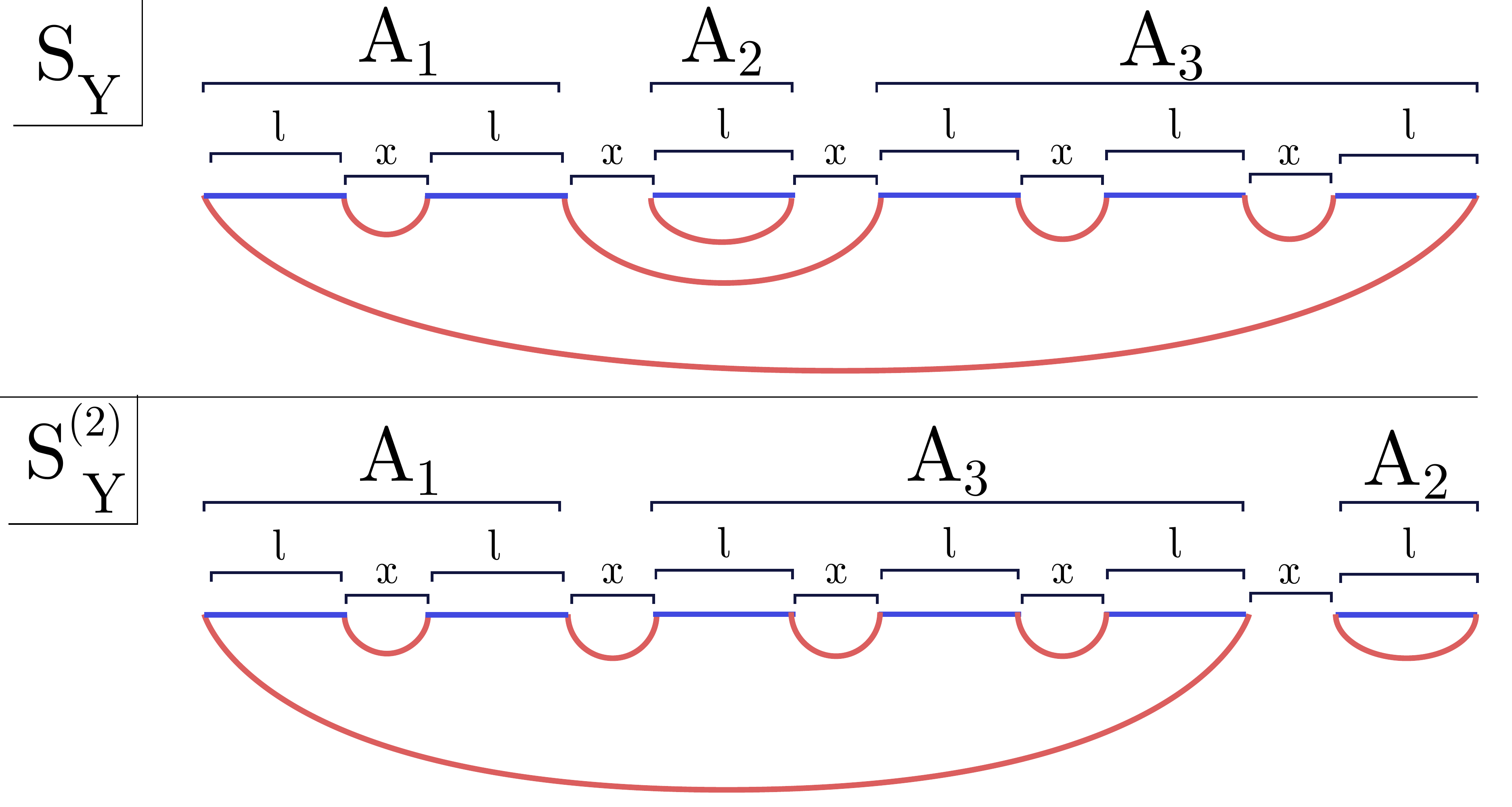}
	\caption{ \textbf{Top:} The class of ``engulfed" bulk minimal surfaces denoted by $S_Y$ where $A_2$ is the ``engulfed" region. \textbf{Bottom:} Permuting the regions $A_2$ and $A_3$. The bulk surface is denoted by $S_{Y}^{(2)}$. We prove that the ``engulfed" bulk minimal surfaces $S_Y$ are not the minimal surfaces. \label{sy1}}
\end{figure}

\begin{figure}
 	\centering
 	\includegraphics[width= 140mm]{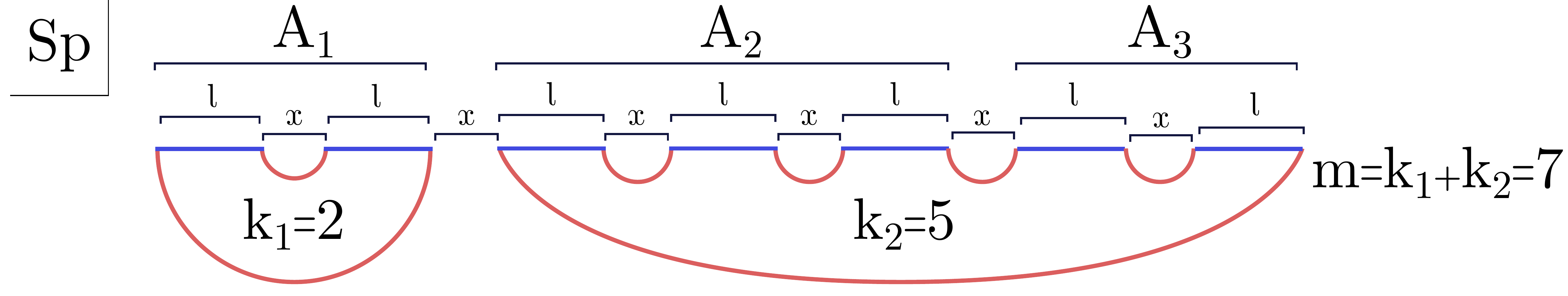}
 	\caption{A class of ``disjoint" minimal surfaces denoted by $S_P$. $S_P$ is composed of two ``disjoint" parts: one of the type $S_B$ with $k_1$ strips, and the other of the type $S_B$ with $k_2$ strips. The plot illustrates this for the case $k_1=2$  and $k_2=5$. The partition to different regions $A_1$, $A_2$ and $A_3$ is also shown. $A_3$ is chosen such that it contains the same number of strips as $A_1$. \label{sp}}
 \end{figure}

 $\bullet$
Consider the class of bulk surfaces (denoted $S_Y$) which have parts that ``engulf" other parts of the bulk surface.
An example of such a surface for the case of 6 strips is shown in Fig.~\ref{sy1}-Top.
For $m$ strips with equal lengths $l$ and equal separations $x$, such surfaces are never the absolute minimal bulk surfaces\footnote{Note that after excluding the classes $S_X$ and $S_Y$, there remain $2^{m-1}$ bulk minimal surfaces. Two of these remaining surfaces are $S_A$ and $S_B$, and the rest are denoted $S_P$ (an example of of such a bulk surface is shown in Fig.~\ref{sp}).}.\\

 \textbf{Proof:}

We will now show that surfaces of the class of $S_Y$ (Fig.~\ref{sy1}-Top) are not the minimal surfaces. We have:
\begin{eqnarray}
\label{eq:syr1}
S_Y = S_{A_1 \cup A_2 \cup A_3} = S_{A_1 \cup A_3} + S_{A_2} 
\end{eqnarray}

Now consider permuting the regions $A_2$ and $A_3$ as in Fig.\ref{sy1}-Bottom. It is clear that $S_Y=S_{A_1 \cup A_2 \cup A_3} = S_{A_1 \cup A_3 \cup A_2}^{(2)}$, where the superscript $(2)$ denotes the system after the permutation. From the monotonicity of the EE we have $S_{A_1 \cup A_3} >S^{(2)}_{A_1 \cup A_3} $, and this inequality is \underline{not} saturated. Therefore Eq.~\ref{eq:syr1} gives:
\begin{eqnarray}
\label{eq:syr5}
S_Y = S_{A_1 \cup A_2 \cup A_3} = S_{A_1 \cup A_3} + S_{A_2} >S^{(2)}_{A_1\cup A_3} +S_{A_2} \geq S_{A_1 \cup A_3 \cup A_2}^{(2)} =S_Y
\end{eqnarray}
where in the second inequality we used subadditivity. We have thus reached a contradiction in Eq.~\ref{eq:syr5}, and therefore $S_Y$ cannot be the bulk minimal surface.

We conjecture (without a proof) that ``engulfed" bulk surfaces are never the (absolute) minimal surfaces even for arbitrary strip lengths $l_i$ and arbitrary separations $x_i$. 
It will be interesting to try and prove this conjecture or to find a counterexample.
\\

$\bullet$
For $m$ strips of equal lengths and equal separations, the ``connected" minimal bulk surface is either $S_A$ or $S_B$, see Fig.~\ref{3stripsmu3}.  
In other words, the ``disjoint" surfaces $S_P$ (exemplified in Fig.~\ref{sp}), are not the absolute minimal surfaces for all values of $x$ and $l$.  \\

 \textbf{Proof:}

Consider $m=k_1+k_2$ equal length and equally separated strips.
Consider also a ``disjoint" bulk surface (with area $S_P$) which is composed of 2 groups of ``joint" surfaces with $k_1$ and $k_2$ strips respectively as shown in Fig.~\ref{sp}. We label  $A_1$, $A_2$ and $A_3$ as in Fig.~\ref{sp}, such that $A_3$ contains $k_1$ strips, as does $A_1$. We know from subadditivity that:
\begin{eqnarray}
\label{eq:saf1}
S_{A_2} +S_{A_3} - S_{A_2 \cup A_3 }\geq 0
\end{eqnarray}
Importantly, this inequality is \underline{not} saturated by the surface $S_P$ of Fig.~\ref{sp} (since if it was saturated then $A_2$ and $A_3$ would be disjoint from each other). 
Since the number of strips in $A_3$ and $A_1$ is the same, and since the strips are of equal lengths and equal separations, we get:
\begin{flalign}
\label{eq:saf2}
S_{A_1}&=S_{A_3}
\nonumber\\
S_{A_2 \cup A_3}&= S_{A_2 \cup A_1}
\end{flalign}

Plugging  Eq.~\ref{eq:saf2} into Eq.~\ref{eq:saf1}, and recalling that the inequality is not saturated, we get:
\begin{eqnarray}
\label{eq:saf4}
S_{A_2} +S_{A_1} - S_{A_2 \cup A_1 } > 0
\end{eqnarray}

For the bulk surface $S_P$ in Fig.~\ref{sp} we have:
\begin{eqnarray}
\label{eq:newr}
\label{eq:stew2} 
S_{A_1 \cup A_2 \cup A_3} = S_{A_2 \cup A_3} + S_{A_1}
\end{eqnarray}

Strong subadditivity (SSA) \cite{Headrick:2007km,Hirata:2006jx}  and Eq.~\ref{eq:newr} give:
\begin{eqnarray}
\label{eq:saf3}
S_{A_1 \cup A_2 \cup A_3} - S_{A_1 \cup A_2 } - S_{A_2 \cup A_3 } + S_{A_2 } = S_{A_2} +S_{A_1} - S_{A_2 \cup A_1 } \leq 0
\end{eqnarray}

We have reached a contradiction between Eqs.~\ref{eq:saf4} and \ref{eq:saf3}. Therefore we exclude the class $S_P$ (consisting of 2 groups of surfaces) of bulk minimal surfaces.\\

Excluding bulk surfaces with 3 or more groups is now a simple task. Consider a ``disjoint" bulk surface which is composed of 3 groups of ``joint" surfaces with $k_1$, $k_2$ and $k_3$ strips respectively. We can apply the 2-group proof to the two groups $k_1$ and $k_2$. As a result we see that there exists a bulk surface with smaller area than the original one, proving what we wanted to show. 

This theorem greatly simplifies the problem of $m$ strips with equal lengths and equal separations, since one has to consider only 2 bulk minimal surfaces (instead of at least $2^{m-1}$ bulk surfaces).

\begin{figure}
	\centering
	
	\begin{minipage}{0.45\textwidth}
		\centering
		\label{sq}
		\includegraphics[height=20mm,width= 90mm]{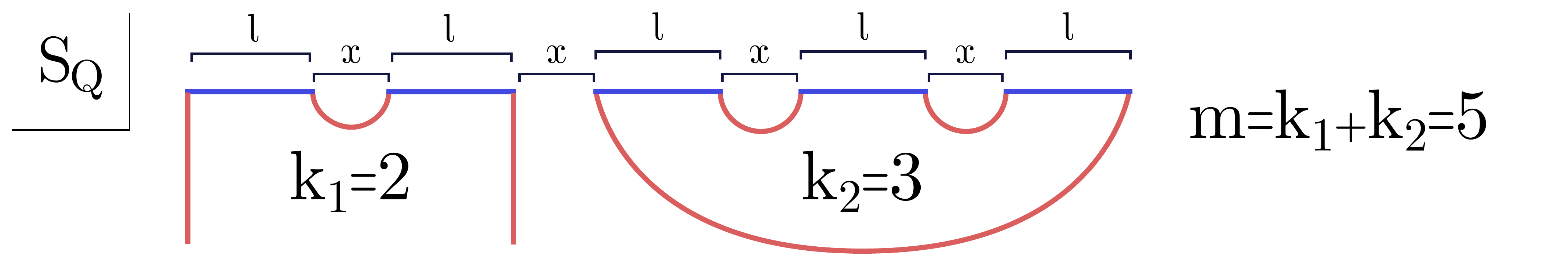}

	\end{minipage}\hfill
	\begin{minipage}{0.45\textwidth}
		\centering
		\label{sr}
		\includegraphics[height=20mm,width= 90mm]{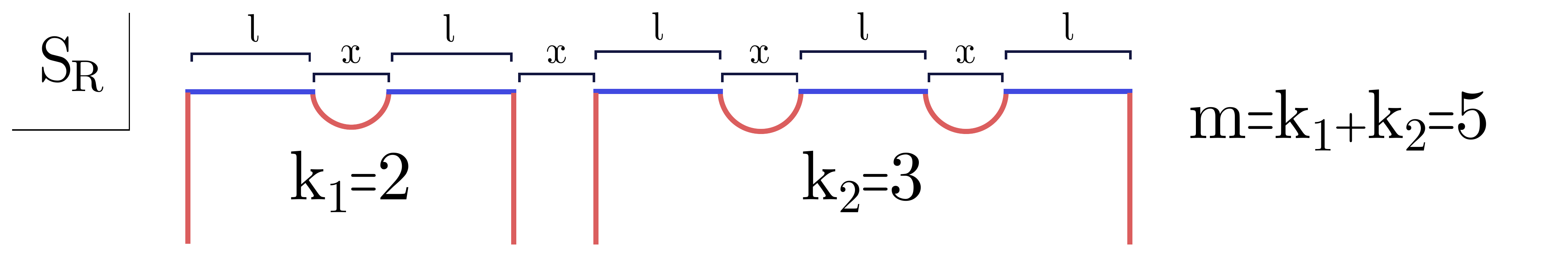}
	\end{minipage}
			\caption{Two classes of ``disjoint" surfaces with a mixture of ``connected" and ``disconnected" parts. The special case of $k_1=2$ and $k_2=3$ is shown. \textbf{Left:} A class of surfaces denoted by $S_Q$. These are comprised of two ``disjoint" parts: an $S_C$ part with $k_1$ strips, and an $S_B$ part with $k_2$ strips. \textbf{Right:} A class of surfaces denoted by $S_R$. These are comprised of two ``disjoint" parts: an $S_C$ part with $k_1$ strips, and an $S_C$ part with $k_2$ strips.}
			\label{fig:disjointdisconnected}

\end{figure}

The above proof also works for surfaces with ``disconnected" parts. Therefore, for equal length strips with equal separations, the bulk minimal surfaces of the classes $S_Q$ or $S_R$ (Fig.~\ref{fig:disjointdisconnected}) are excluded. This brings us to the following result:\\

$\bullet$
For $m$ strips of equal lengths and equal separations the only possible bulk minimal surfaces are~$S_A$, $S_B$, $S_C$, and $S_D$. See~Fig.\ref{3stripsmu3}.\\

As mentioned above, the proof should work for $m$ identical non-strip entangling regions equally separated on a line. Assuming that the topology of the bulk minimal surfaces for this system is similar to that of $m$ strips, there will be 4 possible minimal surfaces $\tilde{S}_A$, $\tilde{S}_B$, $\tilde{S}_C$, $\tilde{S}_D$ with topology similar to $S_A$, $S_B$, $S_C$, and $S_D$.

Additionally, the proof is not limited to holographic entanglement entropy but in principle can be applied to Wilson loops (as long as the latter obey strong subadditivity).

\section{Discussion}
\label{sec:discussion2}

The main goal of this paper was to study HEE of $m$ strips, and transitions between topologically distinct minimal bulk surfaces. We began by analyzing CFTs, and studied the resulting phase diagrams.
For confining backgrounds, the $m$ strip HEE is calculated by new types of ``disconnected" bulk minimal surfaces such as $S_C$ and $S_D$, and the resulting phase diagrams are rich.
 
 Note that there exist other backgrounds (non-confining) for which the 1-strip HEE has a transition between a ``connected" and a ``disconnected" bulk minimal surface. An example is the D3-brane shell model \cite{Bhattacharya:2012mi, Kraus:1998hv}. For such backgrounds with $m$ strips, we expect phase diagrams similar to the confining case, Fig.~\ref{mutualphasediagramAdS5}.
 
The BTZ black hole in global coordinates exhibits the ``entanglement plateau"-like transition in the case of 1 strip. For 2 strips there are 4 possible minimal surfaces. It would be interesting to study the $m$-strip case and also to generalize to higher dimensional black holes.

There is a ``correspondence" between holographic Wilson loops at finite $T$ and HEE for confining backgrounds, and vice versa. Using this, it was shown how our results can be applied (qualitatively) to holographic Wilson loops.

Section~\ref{sec:book12}, contains a perturbative analysis for $m$ strips. A ``positive" perturbation of the CFT, tends to break the ``joint" bulk surfaces into ``disjoint" ones. Conversely, a ``negative" perturbation will tend to join together bulk ``disjoint" surfaces.

Section~\ref{sec:book112} contains a perturbative analysis for $m$ strips where the QFT is not perturbed, but  the length or separation of the strips are. One result, is that the configuration with equal ``inner" separations is a maximum of the HEE with respect to perturbing these ``inner" separations.
A second result, is that enlarging the length of an ``inner" strip, reduces the HEE. This is opposite behavior compared to the effect of enlarging an ``outer" or ``disjoint" strip.

Section~\ref{sec:theorems4} contains a few results which exclude certain classes of bulk minimal surfaces.
In particular, for $m$ strips of equal lengths and equal separations there are only 4 possible bulk minimal surfaces: $S_A$, $S_B$, $S_C$, and $S_D$. This theorem greatly simplifies the problem of $m$ strips with equal lengths and equal separations, since one has to consider only 4 bulk surfaces. Interestingly, it seems that this result is valid also for non-strip regions.

There are several additional questions that follow the analysis performed in this paper.
\begin{itemize}
	\item
		This paper considered strip regions because the translational symmetry effectively reduces the problem of $m$ strips to that of 1 strip.
		It is reasonable to conjecture that for entangling regions which are not strips (for example spheres), the topology of the bulk minimal surfaces will be as for the strip case\footnote{For example \cite{Pakman:2008ui} considers sphere entangling regions for confining backgrounds. They find that there is a ``connected" and a ``disconnected" surface, and a phase transition between the two at a critical value of the radius of the sphere. This is analogous to the strip case of section  \ref{sec:book5}.},  Fig.~\ref{3stripsmu3}. Therefore it is also reasonable to conjecture that the phase diagram in the confining case will qualitatively have the form of Fig.~\ref{mutualphasediagramAdS5}. It might also be interesting to study phase diagrams for concentric spheres. 

	\item
	The dimension of the phase space of most of the systems discussed in this paper is two since we have taken the simplified case of equal strip lengths and equal separation distances. In general for the case of $m$ strips the phase space is of dimension $2m-1$. Analyzing the structure of this multi-dimensional phase space should follow similar procedures as those used in the current simplified case. It is quite probable that determining the general phase space will shed additional light on the considered systems.
	
	\item
	The procedure of  \cite{Calabrese:2009qy} is based on using conformal transformations. One interesting question is if one can generalize this procedure also for computations of the EE of non-conformal and in particular confining backgrounds.

\item In section~\ref{sec:theorems4} it is conjectured that the class of ``engulfed" bulk surfaces denoted $S_Y$ are never the (absolute) minimal surfaces for arbitrary strip lengths and arbitrary separations. It will be interesting to try to prove this conjecture or to find a counterexample.
\end{itemize}

\textbf{Acknowledgments}

We thank Matthew Headrick, Carlos Hoyos, Uri Kol, David Kutasov, Michael Smolkin, Tadashi Takayanagi, Erik Tonni, and  Shimon Yankielowicz for useful discussions. We also thank Avner Gicelter for help with the figures.
This work is partially supported by the Israel Science Foundation (grant 1989/14 ), the  US-Israel bi-national fund (BSF) grant 2012383 and the German Israel bi-national fund GIF grant number I-244-303.7-2013.

\begin{appendix}

\section{Holographic Wilson lines and HEE}
\label{sec:BB5}

\subsection{Holographic Wilson lines in Confining backgrounds}
\label{app:confinement}

In this section we shortly review holographic Wilson lines in confining backgrounds,  see \cite{Maldacena:1998im}, \cite{Kinar:1998vq}.
An important property of confining theories is the area law behavior of the Wilson line. This is equivalent to a linear potential between the quark and anti-quark.

Consider a bulk metric with the following general form:
\begin{eqnarray}
\label{eq:plk2}
ds^2= \alpha_x(U)\big[\beta(U)dU^2+dx^\mu dx_\mu\big]+\alpha_t(U) dt^2+ g^{ij}dy_idy_j
\end{eqnarray}
Where $\alpha_x(U)$, $\alpha_t(U)$ and $\beta(U)$ are functions of the holographic coordinate $U$, $x_\mu$ are the boundary directions ($\mu$ = $1\dots d$) and $y_i$ are internal directions ($i=d+2, \ldots , 10$).

Following \cite{Maldacena:1998im} and \cite{Kinar:1998vq}, the distance between the quark and anti-quark is:
\begin{eqnarray}
\label{dlength321}
l (U^*) = 2 \int _{U^*}^{\infty} \frac{dU \sqrt{\beta} }{ \sqrt{\frac{F^2(U)}{F^2(U^*)}-  1}}
\end{eqnarray}
Where $U^*$ is the lowest point of the string, $U_\infty$ is the UV cutoff, and we defined $F(U) \equiv \sqrt{\alpha_{x}(U)\alpha_{t}(U)}$.

The potential energy between the quark and anti-quark is:
\begin{eqnarray}
\label{denerg321}
V (U^*) = 2 \int _{U^*}^{U_\infty} \frac{dU \sqrt{\beta} F^2(U)}{ \sqrt{F^2(U)-  F^2(U^*)}} -2m_q
\end{eqnarray}
The first term in Eq.~\ref{denerg321} is the bare energy, and the second term is the mass of the quark and anti-quark, which is subtracted in order to renormalize the energy.
The mass of the quarks is obtained from the energy of the two straight strings stretched from $U_\infty$ to $U_0$:
\begin{eqnarray}
m_q= \int _{U_0}^{U_\infty}d U \sqrt{\beta} F(U)
\end{eqnarray}
where $U_0$ is where the space ends.

Linear confinement means that at large $l$ we have:
\begin{equation}
V(l) = F(U_0)\cdot  l + \mathcal{O}(1/l)
\end{equation}

\cite{Kinar:1998vq} showed that a background exhibits linear confinement if one of the two conditions below are satisfied:
\begin{eqnarray}
\label{WLconditions}
\begin{aligned}\bullet \quad & \text{The function } F(U)\:\text{has a minimum.}\\
\bullet \quad & \text{The function } \sqrt{\beta} F(U)\:\text{diverges.}
\end{aligned}
\end{eqnarray}
and also that the tension of the string is non-zero $F(\tilde{U})\neq 0$, where $\tilde U$ is the value at which $F$ is a minimum or the value at which $\sqrt{\beta} F$ diverges.

\begin{figure}[!ht]
	\centering
		\includegraphics[height=5.15cm]{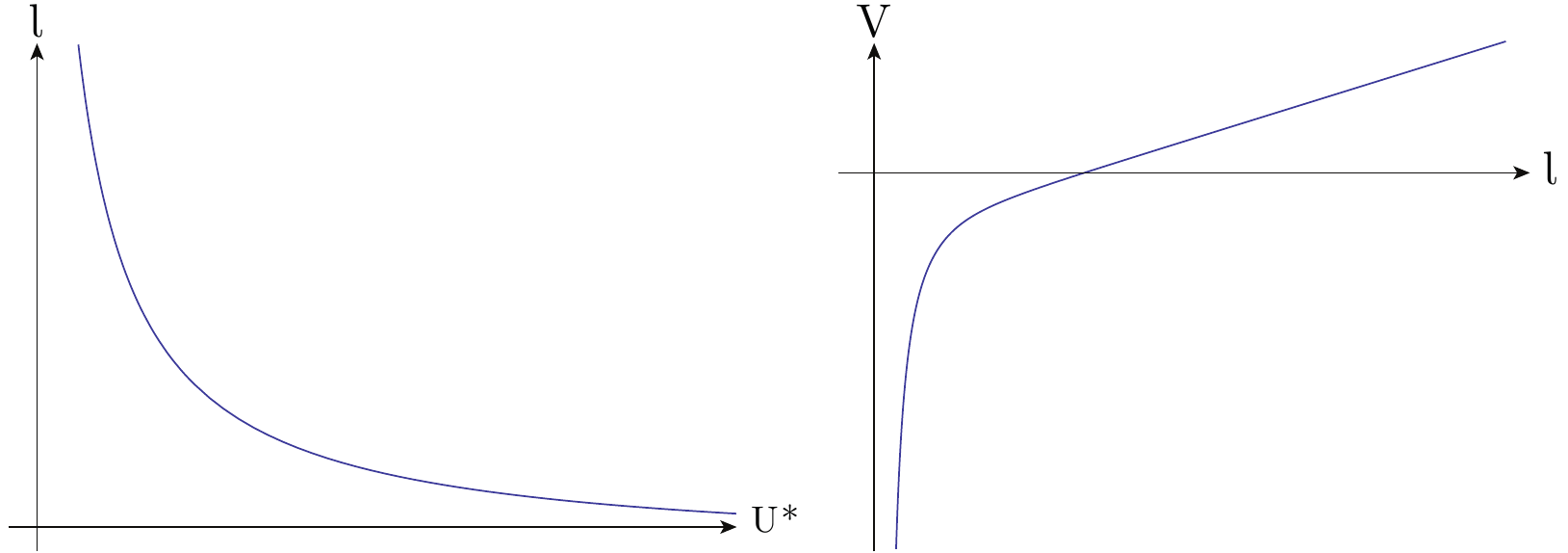}

		\caption{\textbf{Left:} $l(U^*)$ is a monotonically decreasing function of $U^*$. \textbf{ Right:}. $V(l)$. Linear confinement can be seen at large $l$.}
		\label{WL}

\end{figure}

\cite{Kinar:1998vq} proved that the $l(U^*)$ is a monotonically decreasing function of $U^*$ in confining backgrounds. This corresponds to $V(l)$ being a monotonically increasing function of $l$. Fig.~\ref{WL} illustrates the properties of Wilson loops mentioned above.

A confining background is thus defined to be a background for which the holographic rectangular Wilson loop admits such an area law behavior.
The model of a $D4$ brane compactified on a circle \cite{Witten:1998zw} is a prototypical confining background. The non-critical version of this model was studied in \cite{Kuperstein:2004yf}.

\subsection{A correspondence between HEE and holographic Wilson loops}

\label{sec:A}

The calculation of a holographic Wilson loop (HWL) is very similar to that of holographic entanglement entropy (HEE), as both are given by the area of a bulk minimal surface. Let us now obtain the map between the two for the case of a strip.
We consider a metric as in Eq.~\ref{eq:plk2}.

Considering a strip of length $l$, we saw in Eq.~\ref{eq:qq8} that the HEE is obtained by minimizing the following function:
\begin{eqnarray}
\label{eq:qq9}
S=  \frac{\tilde{L}^{d-2}}{4G_N}\int_{-l/2}^{l/2} dx \sqrt{H(U)}\sqrt{1+\beta(\partial_x U)^2}\ \ \ \  .
\end{eqnarray}

where we defined:
\begin{eqnarray}
H(U)\equiv e^{-4\phi}V^2_{int}\alpha_x^{d-1}(U) \ \ \  .
\end{eqnarray}

On the other hand, for holographic Wilson loops we need to minimize the Nambu-Goto action:
\begin{eqnarray}
\label{eq:qq10}
S^{(NG)}= \frac{1}{2\pi \alpha'} \int d\sigma d\tau = \frac{T}{2\pi \alpha'}\int_{-l/2}^{l/2} dx \sqrt{\alpha_x \alpha_t}\sqrt{1+\beta(\partial_x U)^2} .
\end{eqnarray}
Where we chose $\tau=t$ and $\sigma=x$, and $T=\int d\tau$.

So we see that Eq.~\ref{eq:qq9} and Eq.~\ref{eq:qq10} are equal when:
\begin{eqnarray}
\label{eq:673}
H(U) \longrightarrow  \alpha_x \alpha_t
\end{eqnarray} 
So at least formally, we can map a holographic Wilson loop in one geometry to holographic EE in another geometry.
This can be used used in order to find non-trivial properties of entanglement entropy or Wilson loops on the field theory side (see also \cite{Hirata:2008ms}).


\subsection{An example}

An example of this ``correspondence" between HEE and HWL is the following.
There is a similarity between holographic entanglement entropy in confining backgrounds \cite{Klebanov:2007ws,Nishioka:2006gr} and Wilson loops in black hole backgrounds \cite{Brandhuber:1998bs}, and vice versa. 

Schematically:
\begin{align}
S^{(conf.)} (l)\ &\sim \  V^{(BH.)} (l)
\nonumber\\
V^{(conf.)} (l) \ &\sim \  S^{(BH.)} (l)
\end{align}
where $S(l)$ is the EE and $V(l)$ is the quark-antiquark potential. By ``$\sim$", we mean that the two functions qualitatively have a similar shape, as we now show. 

\begin{figure}
	\centering
	\includegraphics[width= 150mm]{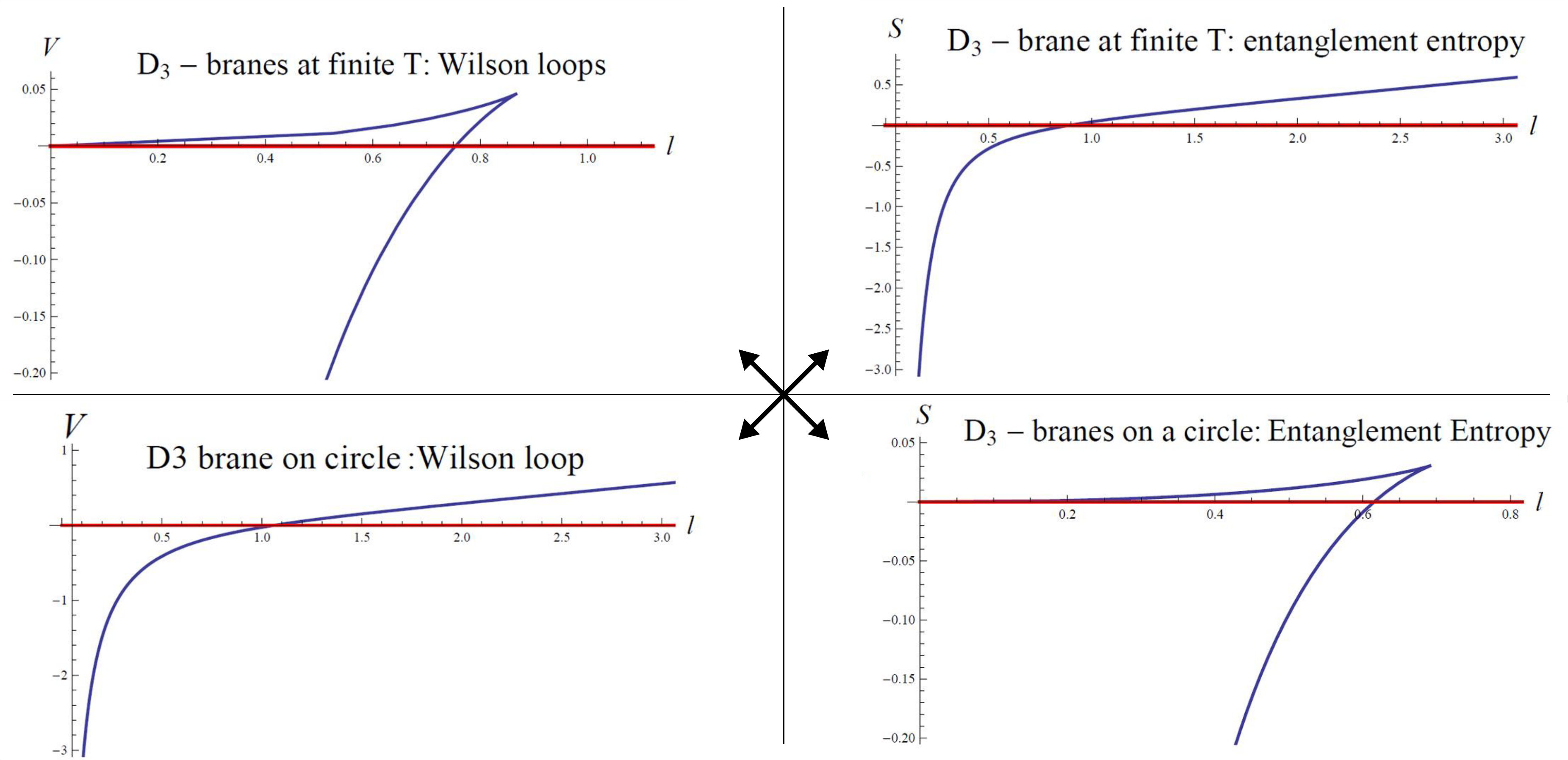}
	\caption{ Showing a ``correspondence" between HEE and HWL. The black arrows show the correspondence. The top-left plot is qualitatively similar to the bottom-right plot. Likewise, the top-right plot is qualitatively similar to the bottom-left plot. \label{corres}}
\end{figure}

To exemplify this ``correspondence" (See Fig.~\ref{corres}), consider the following two backgrounds:\\

\underline{$AdS_5$ compactified on a circle:}
\begin{eqnarray}
ds^2= \Big(\frac{U}{R}\Big)^{2}\Big[ \Big(\frac{R}{U}\Big)^{4}\frac{dU^2}{f(U)} + f(U)dx^2_3 +  dt^2 +  dx_1^2 + dx_2^2 \Big] 
\end{eqnarray}

\underline{$AdS_5$ black hole:}
\begin{eqnarray}
ds^2= \Big(\frac{U}{R}\Big)^{2}\Big[ \Big(\frac{R}{U}\Big)^{4}\frac{dU^2}{f(U)} + f(U)dt^2 + dx_1^2 +  dx_2^2 + dx_3^2 \Big]  
\end{eqnarray}

The two backgrounds are related by $t \leftrightarrow x_3$.

We calculated the HEE and HWL in these two backgrounds, and the result is shown in Fig.~\ref{corres}. The black arrows show the ``correspondence". The WL for the $AdS_5$ BH is qualitatively similar to the EE for $AdS_5$ compactified on a circle. Likewise, the HEE for the $AdS_5$ BH is qualitatively similar to the HWL for $AdS_5$ compactified on a circle.


This qualitative ``correspondence" is true for other confining backgrounds and other $AdS$ black holes.
It can be explicitly seen by looking at the integral expressions for the Wilson loops and entanglement entropy. The ``correspondence" is related to the vanishing of the function $f{(U)}$ at the horizon of a black hole (for the finite $T$ case) and at the tip of the cigar (for the confining case).  

More specifically, the reason it happens is:

1. The Wilson loop picks up the coefficient of time $dt^2$ in the metric, but does not pick up the coefficient of the compact direction $dx_3^2$.

2. The entanglement entropy does not pick up the coefficient of time $dt^2$ in the metric (it is defined at a constant time slice), but does pick up the coefficient of the compact direction $dx_3^2$.

3. The confining metric and the metric of the black hole are related by
exchanging the time direction and the spatial circle:  $dt^2 \leftrightarrow dx_3^2$ :\\

In section~\ref{sec:qq01} we use this correspondence to note that the phase diagrams for Wilson loops of multiple strips, will be qualitatively similar to those of HEE.

\end{appendix}

\bibliographystyle{utphys}

\bibliography{final}

\end{document}